\documentclass[12pt,a4paper]{article}
\pdfoutput=1

\usepackage{jheplike,ifpdf,tocloft,rotating,amsmath,amssymb,mathrsfs,mathtools,subcaption,hyperref,microtype,xcolor}

\widowpenalty=500\clubpenalty=1000\unitlength=1mm\textheight=8.7in\hypersetup{pdftitle={},pdfcreator={},linkcolor=[rgb]{0.15,0.35,0.75},colorlinks=true,citecolor=[rgb]{0.675,0,0.2},urlcolor=[rgb]{0.15,0.35,0.65}}
\let\olditemize\itemize\renewcommand{\itemize}{\vspace{-2pt}\olditemize\setlength{\itemsep}{1pt}\setlength{\parskip}{0pt}\setlength{\parsep}{-0pt}}
\let\oldenumerate\enumerate\renewcommand{\enumerate}{\vspace{-4pt}\oldenumerate\setlength{\itemsep}{1pt}\setlength{\parskip}{0pt}\setlength{\parsep}{0pt}}
\makeatletter\renewcommand\section{\addtocontents{toc}{\protect\addvspace{-2.25\p@}}\@startsection {section}{1}{\z@}{0.5ex \@plus .2ex \@minus 0.2ex}{0.3ex \@plus.1ex\@minus .5ex}{\normalfont\large\bfseries}}
\renewcommand\subsection{\addtocontents{toc}{\protect\addvspace{0.5\p@}}\@startsection {subsection}{1}{\z@}{0.5ex \@plus .2ex \@minus 0.2ex}{0.3ex \@plus.1ex\@minus .5ex}{\normalfont\bfseries}}
\renewcommand\subsubsection{\addtocontents{toc}{\protect\addvspace{-2.5\p@}}\@startsection {subsubsection}{1}{\z@}{0.5ex \@plus .2ex \@minus 0.2ex}{0.3ex \@plus.1ex\@minus .5ex}{\normalfont\bfseries}}
\newtheorem{theorem}{Theorem}

\newcommand{\eq}[1]{\vspace{-0.5pt}\begin{equation}#1\vspace{-0.5pt}\end{equation}}

\newcommand{\fwbox}[2]{\text{\makebox[#1][c]{$\hspace{-150pt}\displaystyle#2\hspace{-150pt}$}}}
\newcommand{\fwboxL}[2]{\text{\makebox[#1][l]{$#2$}}}
\newcommand{\fwboxR}[2]{\text{\makebox[#1][r]{$#2$}}}
\newcommand{\fig}[3]{\raisebox{#1}{\includegraphics[scale=#2]{#3}}}
\newcommand{\bigger}[1]{\raisebox{-0.95pt}{\scalebox{1.25}{$#1$}}}
\newcommand{\mi}{\raisebox{0.75pt}{\scalebox{0.75}{$\hspace{-2pt}\,-\,\hspace{-0.5pt}$}}}
\newcommand{\pl}{\raisebox{0.75pt}{\scalebox{0.75}{$\hspace{-2pt}\,+\,\hspace{-0.5pt}$}}}
\renewcommand{\bar}{\overline}

\renewcommand{\tilde}{\widetilde}
\renewcommand{\phi}{\varphi}

\newcommand{\x}[2]{{\color{black}(}\hspace{-0.85pt}{\color{hblue}#1}\hspace{-0.25pt}{\color{black}|}\hspace{-0.25pt}{\color{hblue}#2}\hspace{-0.85pt}{\color{black})}}
\newcommand{\xHalf}[2]{{\color{black}(}\hspace{-0.85pt}{\color{hblue}#1}\hspace{1.5pt}{\color{black}\nmid}\hspace{1.5pt}{\color{hblue}#2}\hspace{-0.85pt}{\color{black})}}
\renewcommand{\u}[2]{(\hspace{-0.5pt}#1;\hspace{-1.5pt}#2\hspace{-0.5pt})}
\newcommand{\bigproj}[1]{\raisebox{2pt}{\bigger{\big[}}\!#1\!\raisebox{2pt}{\bigger{\big]}}}
\newcommand{\proj}[1]{\raisebox{1pt}{{\big[}}#1\raisebox{1pt}{{\big]}}}
\newcommand{\wpok}{\ensuremath{\mathbb{WP}^{1,\ldots,1,k}}}
\newcommand{\mpl}[2]{G_{\!#1}^{\,\xru_#2}}
\newcommand{\wheelint}{\mathfrak{W}\hspace{-0pt}}
\newcommand{\traintrack}{\mathfrak{T}^{(L)}}
\newcommand{\equivR}{\fwbox{14.5pt}{\hspace{-0pt}\fwboxR{0pt}{\raisebox{0.47pt}{\hspace{1.25pt}:\hspace{-4pt}}}=\fwboxL{0pt}{}}}
\newcommand{\equivL}{\fwbox{14.5pt}{\fwboxR{0pt}{}=\fwboxL{0pt}{\raisebox{0.47pt}{\hspace{-4pt}:\hspace{1.25pt}}}}}

\newcommand{\embd}[1]{{\color{black}|}\hspace{-0.5pt}{\color{hblue}#1}\hspace{-0.75pt}{\color{black})}}
\newcommand{\toyxr}{t}\newcommand{\xru}{u}\newcommand{\xrv}{v}\newcommand{\xrw}{w}

\definecolor{hblue}{rgb}{0,0,0.575}
\definecolor{hred}{rgb}{0.575,0.0,0.225}
\definecolor{hgreen}{rgb}{0.0,0.4,0.2}
\definecolor{dim}{rgb}{0.55,0.55,0.55}

\title{\texorpdfstring{Embedding Feynman Integral (Calabi-Yau) Geometries in Weighted Projective Space\\[-24pt]}{Embedding Feynman Integral (Calabi-Yau) Geometries in Weighted Projective Space}}
\author[a,b]{\vspace{-24pt}Jacob~L.~Bourjaily,}\emailAdd{bourjaily@nbi.ku.dk}
\author[a]{Andrew~J.~McLeod,}\emailAdd{amcleod@nbi.ku.dk}
\author[a]{Cristian~Vergu,}\emailAdd{c.vergu@nbi.ku.dk}
\author[a]{Matthias~Volk,}\emailAdd{mvolk@nbi.ku.dk}
\author[a]{Matt~von~Hippel,}\emailAdd{mvonhippel@nbi.ku.dk}
\author[a]{Matthias~Wilhelm}\emailAdd{matthias.wilhelm@nbi.ku.dk}
\affiliation[a]{Niels Bohr International Academy and Discovery Center, Niels Bohr Institute,\\University of Copenhagen, Blegdamsvej 17, DK-2100, Copenhagen \O, Denmark}
\affiliation[b]{Center for the Fundamental Laws of Nature, Department of Physics,\\ Jefferson Physical Laboratory, Harvard University, Cambridge, MA 02138, USA}
\abstract{%
It has recently been demonstrated that Feynman integrals relevant to a wide range of perturbative quantum field theories involve periods of Calabi-Yau manifolds of arbitrarily large dimension. While the number of Calabi-Yau manifolds of dimension three or higher is considerable (if not infinite), those relevant to most known examples come from a very simple class: degree-$2k$ hypersurfaces in $k$-dimensional weighted projective space $\mathbb{WP}^{1,\ldots,1,k}$. In this work, we describe some of the basic properties of these spaces and identify additional examples of Feynman integrals that give rise to hypersurfaces of this type. Details of these examples at three loops and of illustrations of open questions at four loops are included as ancillary files to this work.}
\preprint{}

\begin{document}
\maketitle\thispagestyle{empty}

%================================================================================================================
%================================================================================================================
% \pagenumbering{roman}
% \clearpage
% \pagenumbering{arabic}

\vspace{\baselineskip}

\vspace{0pt}\section{Introduction and Summary}\label{sec:introduction}\vspace{0pt}

Recent years have seen the development of a rich interplay between number theory, algebraic geometry, and the study of perturbative scattering amplitudes in quantum field theory. Even for what is arguably the simplest class of amplitudes---those that can be expressed in terms of multiple polylogarithms~\cite{Chen,G91b,Goncharov:1998kja,Remiddi:1999ew,Borwein:1999js,Moch:2001zr}---a great deal of conceptual and computational progress has been made~\cite{Goncharov:2010jf,CaronHuot:2011ky,Duhr:2012fh,Dixon:2013eka,Dixon:2014voa,Dixon:2014iba,Drummond:2014ffa,Dixon:2015iva,Panzer:2016snt,Caron-Huot:2016owq,Dixon:2016apl,Dixon:2016nkn,Almelid:2017qju,Schnetz:2017bko,Caron-Huot:2018dsv,Drummond:2018caf,DelDuca:2018raq,Abreu:2018aqd,Chicherin:2018yne,Chicherin:2019xeg,Abreu:2019rpt,Caron-Huot:2019vjl,Caron-Huot:2019bsq} by harnessing the geometric (or motivic) structures with which these functions are endowed when viewed as iterated integrals on the moduli space of the Riemann sphere with marked points~\cite{G91b,Goncharov:1998kja,Goncharov:2001iea,Goncharov:2005sla,Brown:2009qja,Brown:2011ik,Brown1102.1312,2011arXiv1101.4497D,Duhr:2011zq,Schnetz:2013hqa,Brown:2015fyf}.

Slightly more complicated amplitudes can be described in terms of elliptic multiple polylogarithms, which can be understood as iterated integrals over the (moduli space of the) torus. This class of functions  has been the focus of a great deal of recent work and is now also under reasonably good theoretical control (in part based on an understanding of modular forms)~\cite{Laporta:2004rb,MullerStach:2012az,brown2011multiple,Bloch:2013tra,Adams:2013kgc,Adams:2014vja,Adams:2015gva,Adams:2015ydq,Adams:2016xah,Adams:2017ejb,Adams:2017tga,Bogner:2017vim,Broedel:2017kkb,Broedel:2017siw,Adams:2018yfj,Broedel:2018iwv,Adams:2018bsn,Broedel:2018qkq,Adams:2018kez,Honemann:2018mrb,Bogner:2019lfa,Broedel:2019hyg,Broedel:2019kmn}.

In general, one expects increasingly complicated classes of integrals to appear in scattering amplitudes at higher perturbative orders, corresponding to integrals over manifolds with higher dimension and/or genus. Even for amplitudes known or expected to be polylogarithmic, this feature may be impossible to realize while preserving locality (see for example ref.~\cite{Bourjaily:2015bpz} and the examples discussed in ref.~\cite{Bourjaily:2019iqr}). A general understanding of the types of integrals that can show up is currently lacking. However, in a surprisingly large number of cases, it has been observed that these manifolds are Calabi-Yau \cite{Brown:2009ta,Brown:2010bw,Bourjaily:2018yfy,Bourjaily:2018ycu,Festi:2018qip,Broedel:2019kmn,Besier:2019hqd,mirrors_and_sunsets}.

Even at dimensions as low as three or four, large numbers of Calabi-Yau manifolds are known to exist---having been constructed and studied, in part, because of their role in string compactifications (see e.g.~refs.~\cite{Braun:2010vc,Anderson:2011ns,Anderson:2012yf,Gray:2013mja}). One may wonder if a similarly vast number of geometries are relevant to Feynman integrals in perturbative quantum field theories. The answer seems to be \emph{no}. Indeed, all the examples identified in ref.\ \cite{Brown:2009ta} and the entire class of `maximally rigid, marginal' integrals described in ref.\ \cite{Bourjaily:2018yfy} are members of a special family: they are given as codimension-one (degree-$2k$) hypersurfaces in the $k$-dimensional weighted projective space $\mathbb{WP}^{1,\ldots,1,k}$. This motivates us to better understand this family of Calabi-Yau manifolds and explore the consequences of their geometry for physics.

The coefficients of the polynomials that define these hypersurfaces are functions of kinematic data. Virtually all known examples involve (highly) \emph{singular} Calabi-Yau hypersurfaces, and there is little doubt that these singularities will play a significant role in our understanding of these Feynman integrals. But in this work we mostly set these bigger questions aside and discuss the geometry of the smooth case---obtainable, in general, by sufficiently `regularizing' complex structure deformations. 
This regularization makes it possible for us to compute various
topological quantities, such as Hodge numbers.  The reader may wonder
what is the significance of these quantities.  While a complete answer
is not available at this point, we believe it is likely that the Hodge
numbers will account for part of the contribution to the dimension of
integral bases in terms of which integrals sharing a given topology
decompose.\\

This work is organized as follows. In section~\ref{sec:weighted-projective-space}, we review some basic algebraic and differential-geometric aspects of these particular Calabi-Yau geometries. In particular, we discuss their Dolbeault cohomology groups $H^{p,q}$ and how to compute the associated Hodge numbers $h^{p,q}$, and discuss the Euler characteristics of (smooth) Calabi-Yau hypersurfaces of $\mathbb{WP}^{1,\ldots,1,k}$. We also review the construction of canonical holomorphic forms (unique up to an overall scaling), and discuss how the integral of this form over various cycles defines the independent periods of the hypersurface (which in some sense characterize its geometry).

We study these aspects of Calabi-Yau hypersurfaces in $\mathbb{WP}^{1,\ldots,1,k}$ with the general expectation that the integral geometries appearing in Feynman diagrams can be found to encode some of the physics of these diagrams. Characterizing these geometries is a first and necessary step for identifying such connections. We also expect these geometries to be relevant to the development of technology for representing these Feynman integrals in terms of iterated integrals. For instance, periods play an important role in the definition of elliptic multiple polylogarithms~\cite{Broedel:2017kkb,Broedel:2017siw,Broedel:2018iwv,Broedel:2018qkq} and are required to bring differential equations into $\epsilon$-canonical form~\cite{Adams:2018yfj}. However, for general Calabi-Yau $(k{-}1)$-folds beyond the elliptic case ($k=2$), the calculation of these periods still poses a challenging problem. (But see ref.~\cite{Candelas:1990rm} for an example where it has been done for the quintic Calabi-Yau hypersurface in $\mathbb{P}^4$.)
Additionally, a connection between the dimension of certain cohomology groups and numbers of master integrals has recently been established using intersection theory~\cite{Mizera:2017rqa,Mastrolia:2018uzb,Frellesvig:2019uqt}.

After this geometric primer, we go on in section~\ref{sec:three-loop} to describe in detail two examples of Calabi-Yau geometries relevant to massless, four-dimensional planar theories at three loops. Unlike the analysis in refs.~\cite{Brown:2009ta,Bourjaily:2018yfy}, which identified these geometries using direct integration, we here identify such hypersurfaces by taking sequences of residues (as done in ref.~\cite{Bourjaily:2018ycu}). We do this by first deriving manifestly dual-conformal-invariant six-fold representations of these integrals using loop-by-loop Feynman parametrization \cite{Bourjaily:2019jrk,Bourjaily:2017bsb,Bourjaily:2018ycu,Bourjaily:2018aeq}. As both integrals contribute to planar maximally supersymmetric Yang-Mills theory, we expect all six remaining integrations to be transcendental; therefore, each residue mimics a polylogarithmic integration.\footnote{This follows from the expectation that three-loop integrals will evaluate to functions with uniform transcendental weight six. Even though the notion of transcendental weight is not established beyond the case of polylogarithms, both integrals degenerate to weight-six polylogarithms in known limits.} In both integrals, Calabi-Yau hypersurfaces appear in the denominator when no more residues can be taken.

\begin{figure}
 \centering
 \begin{subfigure}{0.45\textwidth}
 \centering
  \includegraphics{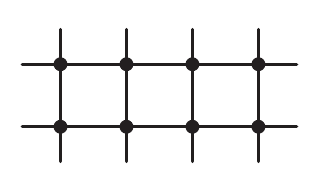}
  \caption{\,}
  \label{subfig: three-loop-traintrack}
 \end{subfigure}
 \begin{subfigure}{0.45\textwidth}
 \centering
  \fig{-24.65pt}{1}{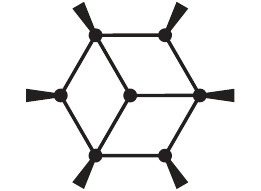}
  \caption{\vspace{-0pt}}
  \label{subfig: three-loop-wheel}
 \end{subfigure}\vspace{-8pt}
 \caption{The three-loop traintrack (a) and wheel (b) integrals.}
 \label{fig:planarCYs}\vspace{-18pt}
\end{figure}

The first integral we study in this way is the three-loop traintrack (or triple-box) integral shown in figure \ref{subfig: three-loop-traintrack}, which has already been identified as a K3 surface by several of the authors \cite{Bourjaily:2018ycu}. We show here how to realize it as a hypersurface in $\mathbb{WP}^{1,1,1,3}$.
The second integral is the three-loop wheel shown in figure \ref{subfig: three-loop-wheel}, which involves a hypersurface in $\mathbb{WP}^{1,1,1,1,4}$. While the general three-loop wheel depends on nine kinematic variables,
we also study several of its interesting kinematic limits, some of which we evaluate in terms of polylogarithms. Moreover, we show that the three-loop wheel permits a toy model similar to that of the elliptic double-box \cite{Bourjaily:2017bsb}, which has only three parameters while still involving a Calabi-Yau threefold.

The three-loop traintrack and wheel integrals are the minimal representatives (in terms of loop order and particle multiplicity) of massless planar topologies that contain these Calabi-Yau geometries.
They occur in massless $\phi^4$ theory (in the case of the wheel, as a dual graph), the planar limit of maximally supersymmetric Yang-Mills theory, and integrable conformal fishnet models \cite{Gurdogan:2015csr,Sieg:2016vap,Grabner:2017pgm}, as well as in more general four-dimensional massless theories via generalized unitarity \cite{Bern:1994zx,Bern:1994cg,Bern:1997sc,Britto:2004nc,Bern:2007ct,Bourjaily:2017wjl,integrandBases}. For this reason, they merit focused investigation. While the present work inaugurates this study, it offers only a coarse analysis of the involved Calabi-Yau geometries. A more refined analysis, including e.g.\ Picard ranks, has been possible for some integrals containing K3 surfaces \cite{Brown:2010bw,Festi:2018qip,Bloch:2014qca,Bloch:2016izu,Besier:2019hqd,mirrors_and_sunsets}, for instance, using differential equations.
It would be important to analyze the Calabi-Yau surfaces identified here and in refs.~\cite{Brown:2009ta,Bourjaily:2018yfy,Bourjaily:2018ycu} in a similar way, although these cases will be more difficult due to the larger number of kinematic variables.

We conclude in section \ref{sec:open-problems} by highlighting open problems at four loops and beyond. In addition to discussing some of the broader questions that remain to be answered regarding the appearance of higher-dimensional varieties in Feynman integrals (and the technology required to cope with them), we consider the four-loop traintrack and wheel integrals. Intriguingly, we are not able to identify either of these example as a Calabi-Yau hypersurface in $\mathbb{WP}^{1,\dots,1,k}$.

In appendices~\ref{sec:desing_cplx_structure} and~\ref{sec:hodge_numbers}, we provide more background on the desingularization of hypersurfaces in weighted projective space and the computation of Hodge numbers and Euler characteristics. In appendix~\ref{appendix:derivation_of_the_wheel} we review loop-by-loop Feynman parametrization~\cite{Bourjaily:2019jrk,Bourjaily:2017bsb,Bourjaily:2018ycu,Bourjaily:2018aeq} and derive a manifestly dual-conformal six-fold representation of the three-loop wheel integral, and in appendix~\ref{appendix:derivation_of_the_wheel_four} we derive a dual-conformal nine-fold representation of the four-loop wheel. In the latter case, we also describe several interesting kinematic limits and toy models. In an ancillary file, we include the details of these examples, as well as the equations defining the hypersurfaces obtained.

%\newpage
\section{Calabi-Yau Hypersurfaces in \texorpdfstring{$\mathbb{WP}^{1,\ldots,1,k}$}{WP(1,...,1,k)}}\label{sec:weighted-projective-space}%

In this section, we characterize the $k$-dimensional weighted projective space $\wpok$, which involves $k$ coordinates of weight 1 and a single coordinate of weight $k$.
This space can be defined as the quotient of $\mathbb{C}^{k+1} \setminus \{0\}$ by the equivalence relation
\begin{align}
\label{eq:equivalence-relation-wpok}
  (x_1, \ldots, x_k,y) \sim (\lambda x_1, \ldots, \lambda x_{k}, \lambda^k y)\,.
\end{align}
Here, $\lambda \in \mathbb{C}^\star$ denotes a non-zero complex number and $(x_1, \ldots, x_k, y)$ are referred to as homogeneous coordinates on $\wpok$.

We will be interested in defining an algebraic hypersurface embedded into $\wpok$ as the zero-locus of a polynomial $Q$ in the homogeneous coordinates.
Of course, such a polynomial relation has to be consistent with the equivalence relation~\eqref{eq:equivalence-relation-wpok}.
In unweighted projective space, this would correspond to the requirement that the polynomial be homogeneous. Analogously, in weighted projective space, the total weight of each monomial must be the same; this number is called the (overall) degree of the polynomial.

One can show (see for example ref.~\cite{Hubsch:1992nu}) that the zero-locus of any single polynomial in the coordinates of a weighted projective space defines a codimension-one Calabi-Yau hypersurface if the overall degree of the polynomial equals the sum of the weights of the weighted projective space.
In the case of $\wpok$, a Calabi-Yau hypersurface can thus be defined by a polynomial $Q$ of degree $(\sum_{i=1}^{k} 1) + k = 2k$, which has the most general form 
\begin{align}
  \label{eq:defining-eq-hypersurface-wpok}
  Q(x_1, \ldots, x_k, y) = \sum_{\substack{(\vec{\alpha},\beta) \in \mathbb{N}_0^{k+1}\\ |\vec{\alpha}|+ k \beta = 2k}} c_{\vec{\alpha},\beta} \, \prod_{i=1}^k x_i^{\alpha_i}y^\beta  \, ,
\end{align}
where $\alpha_i$ denotes the $i^\text{th}$ component of $\vec{\alpha}$ and $|\vec{\alpha}| \equivR \sum_i \alpha_i$. The coefficients $c_{\vec{\alpha},\beta} \in \mathbb{C}$ are complex numbers, and can in general depend on additional parameters (which for us will be kinematics). However, these coefficients are only defined up to $\wpok$ coordinate transformations. In particular, we can rescale all coordinates using the equivalence relation \eqref{eq:equivalence-relation-wpok} to set $c_{\vec{0},2}\to1$, and additionally shift $y$ by a degree-$k$ polynomial in the $x_i$ to eliminate the terms linear in $y$ (thereby setting $c_{\vec{\alpha},1}\to0$).
This brings $Q$ into the form 
\begin{equation}
\label{eq:defining-eq-hypersurface-wpok-simplified}
 Q(x_1, \ldots, x_k, y)=y^2-P(x_1, \ldots, x_k).
\end{equation}
Finally, we can act with a $GL(k)$ transformation on the $x_i$. This can be used to eliminate $k^2$ of the $\binom{3k-1}{k-1}$ possible monomials in $P$.
The remaining $\binom{3k-1}{k-1}-k^2$ coefficients yield distinct hypersurfaces, which are usually parametrized by $\binom{3k-1}{k-1}-k^2$ complex structure moduli.
We should emphasize that hypersurfaces taking the form~\eqref{eq:defining-eq-hypersurface-wpok} may be singular for some values of the coefficients.  For generic coefficients, they are however smooth (see the discussion in appendix~\ref{sec:desing_cplx_structure}).

We now consider a Calabi-Yau manifold $X$ embedded as a codimension-one hypersurface in $\wpok$ and study the forms on $X$.
Since $X$ is a complex manifold, any $m$-form on $X$ can be decomposed into a sum of forms with $p$ holomorphic and $q$ antiholomorphic pieces such that $p + q = m$.
Moreover, the exterior derivative decomposes as $\operatorname{d} = \partial + \bar{\partial}$.
In analogy with de Rham cohomology, one can then define the Dolbeault cohomology groups $H^{p,q}(X)$ as the cohomology groups of $\bar{\partial}$.
The dimensions of the Dolbeault cohomology groups are known as Hodge numbers, $h^{p,q}(X) \equivR \operatorname{dim}(H^{p,q}(X))$.
Moreover, the dimensions of the de Rham cohomology groups $h^m$ are given by $h^m = \sum_{p + q = m} h^{p,q}$.
Recall that via Poincar\'e duality and de Rham's theorem, $h^m$ are exactly the Betti numbers, which count the numbers of independent $m$-cycles on $X$.

In an $n$-dimensional complex manifold in which $p$ and $q$ run from $0$ to $n$, one might na\"{i}vely expect $(n+1)^2$ different Hodge numbers.
However, due to various symmetries many of these numbers are not independent.
For example, in the case of a Calabi-Yau threefold ($k = 4$), $h^{1,1}$ and $h^{2,1}$ fix the values of all other Hodge numbers.

In general, the computation of the Hodge numbers of a complex manifold poses a difficult problem.\footnote{For smooth varieties, this can be achieved in general by
Gr\"obner bases computations.  We thank the referee for pointing this out to us.}
In the case of Calabi-Yau hypersurfaces embedded in toric varieties---of which weighted projective space is an example---the mirror-symmetry construction due to Batyrev~\cite{MR1269718} provides a framework to compute (some of the) Hodge numbers from purely combinatorial data. (For more pedagogical introductions on this topic see, for instance, refs.~\cite{Danilov_1978,MR2003030}.)
In short, one associates to a defining polynomial (such as $Q$ in eqn.~\eqref{eq:defining-eq-hypersurface-wpok}) a pair of dual polytopes $(\Delta, \Delta^\star)$.
The polytope $\Delta$ is called the Newton polytope and its vertices are given by the (shifted) vectors of exponents of the polynomial.
(Note that the vertices of $\Delta$ therefore lie in an integer lattice.)
One can show that in terms of $(\Delta, \Delta^\star)$ the Calabi-Yau condition becomes the statement that the dual polytope $\Delta^\star$ only has integer vertices and that both polytopes contain only the origin as an interior lattice point.
Some of the Hodge numbers can then be computed from the polytopes as
\eq{
\label{eq:h1n-batyrev}
\begin{split}
\hspace{-20pt}h^{1,n} =\,&
  \delta_{1,n} \Big[ \ell(\Delta^\star) - (d + 1) - \sum_{\fwbox{0pt}{\operatorname{codim} \theta^\star\!\!= 1}} \ell_{\mathrm{int}}(\theta^\star) \Big]+
  \delta_{d-2,n} \Big[ \ell(\Delta) - (d + 1) - \sum_{\fwbox{0pt}{\operatorname{codim} \theta = 1}} \ell_{\mathrm{int}}(\theta) \Big]\hspace{-26pt}\\
  & +
  \sum_{\fwbox{0pt}{\operatorname{codim}\theta^\star\!\!= n + 1}} \ell_{\mathrm{int}}(\theta^\star) \ell_{\mathrm{int}}(\theta).
\end{split}
}
Here $\ell$ and $\ell_{\mathrm{int}}$ count the total and interior lattice points of a polytope, respectively, and the sums run over faces of $\Delta$ and $\Delta^\star$, denoted by $\theta$ and $\theta^\star$, with the given codimension.
Note that Batyrev's framework explicitly excludes the case of K3 surfaces ($k=3$).

This construction can be generalized to so-called complete intersection Calabi-Yaus (CICYs) embedded into a toric variety and one can obtain more Hodge numbers as the expansion coefficients of a two-variable generating function known as stringy $E$-function~\cite{Batyrev:1995ca,Kreuzer:2001fu},
\begin{align}
  \label{eq:stringy-e-fct}
  E(u, v) = \sum_{p,q} (-1)^{p+q} h^{p,q} u^p v^q.
\end{align}
The construction of the function $E$ relies on a generalization of the reflexive polytope criterion outlined above by so-called nef-partitions.
The function has been implemented in \texttt{PALP}~\cite{Kreuzer:2002uu}, which is available from \texttt{SageMath}~\cite{sagemath}.

For the case of a Calabi-Yau hypersurface in $\wpok$, the Newton polytope and its dual take a relatively simple form and allow us to compute $h^{1,j}$ from eqn.~\eqref{eq:h1n-batyrev}: for any $k\geq 4$, we find
\begin{align}
  \label{eq:h1n-in-wpok}
  h^{1,j} =
  \begin{cases}
   \binom{3k-1}{k-1}-k^2 & j = k - 2, \\
    1                                                                 & j = 1, \\
    0                                                                 & \text{otherwise.}
  \end{cases}
\end{align}
The non-trivial Hodge number $h^{1,k-2}$ counts the complex structure moduli discussed above% (see also appendix~\ref{sec:cplx_moduli})
, while $h^{1,1}$ counts the single K\"ahler structure modulus.
We have moreover verified this formula by comparing to the stringy $E$-function implemented in \texttt{PALP}~\cite{Kreuzer:2002uu}, which also computes the remaining Hodge numbers.

For the elliptic curve ($k=2$), the Hodge numbers are well-known to be
\eq{\begin{array}{@{}c@{}}\fwboxL{30pt}{h^{0,0}}\\[-1pt]
\fwboxL{30pt}{h^{1,0}}\fwboxL{30pt}{h^{0,1}}\\[-1pt]
\fwboxL{30pt}{h^{1,1}}
\end{array}=\begin{array}{@{}c@{}}\fwbox{26pt}{1}\\[-2pt]
\fwbox{26pt}{1}\fwbox{26pt}{1}\\[-2pt]
\fwbox{26pt}{1}
\end{array}.}
As already mentioned, the case of the K3 surface, $k=3$, is excluded in the general framework above.
Here, in addition to the $\binom{6 + 2}{2} - 9 = 19$ complex structure moduli, the K\"ahler structure modulus contributes to $h^{1,1}$, allowing us to obtain the well-known result
\eq{\begin{array}{@{}c@{}}\fwboxL{30pt}{h^{0,0}}\\[-1pt]
\fwboxL{30pt}{h^{1,0}}\fwboxL{30pt}{h^{0,1}}\\[-1pt]
\fwboxL{30pt}{h^{2,0}}\fwboxL{30pt}{h^{1,1}}\fwboxL{30pt}{h^{0,2}}\\[-1pt]
\fwboxL{30pt}{h^{2,1}}\fwboxL{30pt}{h^{1,2}}\\[-1pt]
\fwboxL{30pt}{h^{2,2}}
\end{array}=\begin{array}{@{}c@{}}\fwbox{24pt}{1}\\[-2pt]
\fwbox{24pt}{{\color{dim}0}}\fwbox{24pt}{{\color{dim}0}}\\[-2pt]
\fwbox{24pt}{1}\fwbox{24pt}{20}\fwbox{24pt}{1}\\[-2pt]
\fwbox{24pt}{{\color{dim}0}}\fwbox{24pt}{{\color{dim}0}}\\[-2pt]
\fwbox{24pt}{1}
\end{array}.\label{eq:K3_hodge_diamond}}
For higher $k$, we find the following patterns of Hodge numbers:
\begin{itemize}
\item Calabi-Yau threefold, $k = 4$:
\eq{\begin{array}{@{}c@{}}\fwboxL{30pt}{h^{0,0}}\\[-1pt]
\fwboxL{30pt}{h^{1,0}}\fwboxL{30pt}{h^{0,1}}\\[-1pt]
\fwboxL{30pt}{h^{2,0}}\fwboxL{30pt}{h^{1,1}}\fwboxL{30pt}{h^{0,2}}\\[-1pt]
\fwboxL{30pt}{h^{3,0}}\fwboxL{30pt}{h^{2,1}}\fwboxL{30pt}{h^{1,2}}\fwboxL{30pt}{h^{0,3}}\\[-1pt]
\fwboxL{30pt}{h^{3,1}}\fwboxL{30pt}{h^{2,2}}\fwboxL{30pt}{h^{2,3}}\\[-1pt]
\fwboxL{30pt}{h^{3,2}}\fwboxL{30pt}{h^{2,3}}\\[-1pt]
\fwboxL{30pt}{h^{3,3}}
\end{array}=\begin{array}{@{}c@{}}\fwbox{24pt}{1}\\[-2pt]
\fwbox{24pt}{{\color{dim}0}}\fwbox{24pt}{{\color{dim}0}}\\[-2pt]
\fwbox{24pt}{{\color{dim}0}}\fwbox{24pt}{1}\fwbox{24pt}{{\color{dim}0}}\\[-2pt]
\fwbox{24pt}{1}\fwbox{24pt}{149}\fwbox{24pt}{149}\fwbox{24pt}{1}\\[-2pt]
\fwbox{24pt}{{\color{dim}0}}\fwbox{24pt}{1}\fwbox{24pt}{{\color{dim}0}}\\[-2pt]
\fwbox{24pt}{{\color{dim}0}}\fwbox{24pt}{{\color{dim}0}}\\[-2pt]
\fwbox{24pt}{1}
\end{array}.}
\item Calabi-Yau fourfold, $k = 5$:
\eq{\begin{array}{@{}c@{}}\fwboxL{30pt}{h^{0,0}}\\[-4pt]
\fwboxL{30pt}{h^{1,0}}\fwboxL{30pt}{h^{0,1}}\\[-4pt]
\fwboxL{30pt}{h^{2,0}}\fwboxL{30pt}{h^{1,1}}\fwboxL{30pt}{h^{0,2}}\\[-4pt]
\fwboxL{30pt}{h^{3,0}}\fwboxL{30pt}{h^{2,1}}\fwboxL{30pt}{h^{1,2}}\fwboxL{30pt}{h^{0,3}}\\[-4pt]
\fwboxL{30pt}{h^{4,0}}\fwboxL{30pt}{h^{3,1}}\fwboxL{30pt}{h^{2,2}}\fwboxL{30pt}{h^{1,3}}\fwboxL{30pt}{h^{0,4}}\\[-4pt]
\fwboxL{30pt}{h^{4,1}}\fwboxL{30pt}{h^{3,2}}\fwboxL{30pt}{h^{1,3}}\fwboxL{30pt}{h^{1,4}}\\[-4pt]
\fwboxL{30pt}{h^{4,2}}\fwboxL{30pt}{h^{3,3}}\fwboxL{30pt}{h^{2,4}}\\[-4pt]
\fwboxL{30pt}{h^{4,3}}\fwboxL{30pt}{h^{3,4}}\\[-4pt]
\fwboxL{30pt}{h^{4,4}}
\end{array}=\begin{array}{@{}c@{}}\fwbox{30pt}{1}\\[-4pt]
\fwbox{30pt}{{\color{dim}0}}\fwbox{30pt}{{\color{dim}0}}\\[-4pt]
\fwbox{30pt}{{\color{dim}0}}\fwbox{30pt}{1}\fwbox{30pt}{{\color{dim}0}}\\[-4pt]
\fwbox{30pt}{{\color{dim}0}}\fwbox{30pt}{{\color{dim}0}}\fwbox{30pt}{{\color{dim}0}}\fwbox{30pt}{{\color{dim}0}}\\[-4pt]
\fwbox{30pt}{1}\fwbox{30pt}{976}\fwbox{30pt}{3952}\fwbox{30pt}{976}\fwbox{30pt}{1}\\[-4pt]
\fwbox{30pt}{{\color{dim}0}}\fwbox{30pt}{{\color{dim}0}}\fwbox{30pt}{{\color{dim}0}}\fwbox{30pt}{{\color{dim}0}}\\[-4pt]
\fwbox{30pt}{{\color{dim}0}}\fwbox{30pt}{1}\fwbox{30pt}{{\color{dim}0}}\\[-4pt]
\fwbox{30pt}{{\color{dim}0}}\fwbox{30pt}{{\color{dim}{\color{dim}0}}}\\[-4pt]
\fwbox{28pt}{1}
\end{array}.}
\item Calabi-Yau fivefold, $k = 6$:
\eq{\begin{array}{@{}c@{}}\fwbox{40pt}{1}\\[-4pt]
\fwbox{40pt}{{\color{dim}0}}\fwbox{40pt}{{\color{dim}0}}\\[-4pt]
\fwbox{40pt}{{\color{dim}0}}\fwbox{40pt}{1}\fwbox{40pt}{{\color{dim}0}}\\[-4pt]
\fwbox{40pt}{{\color{dim}0}}\fwbox{40pt}{{\color{dim}0}}\fwbox{40pt}{{\color{dim}0}}\fwbox{40pt}{{\color{dim}0}}\\[-4pt]
\fwbox{40pt}{{\color{dim}0}}\fwbox{40pt}{{\color{dim}0}}\fwbox{40pt}{1}\fwbox{40pt}{{\color{dim}0}}\fwbox{40pt}{{\color{dim}0}}\\[-4pt]
\fwbox{40pt}{1}\fwbox{40pt}{6152}\fwbox{40pt}{67662}\fwbox{40pt}{67662}\fwbox{40pt}{6152}\fwbox{40pt}{1}\\[-4pt]
\fwbox{40pt}{{\color{dim}0}}\fwbox{40pt}{{\color{dim}0}}\fwbox{40pt}{1}\fwbox{40pt}{{\color{dim}0}}\fwbox{40pt}{{\color{dim}0}}\\[-4pt]
\fwbox{40pt}{{\color{dim}0}}\fwbox{40pt}{{\color{dim}0}}\fwbox{40pt}{{\color{dim}0}}\fwbox{40pt}{{\color{dim}0}}\\[-4pt]
\fwbox{40pt}{{\color{dim}0}}\fwbox{40pt}{1}\fwbox{40pt}{{\color{dim}0}}\\[-4pt]
\fwbox{40pt}{{\color{dim}0}}\fwbox{40pt}{{\color{dim}0}}\\[-4pt]
\fwbox{40pt}{1}
\end{array}.}
\item Calabi-Yau sixfold, $k = 7$:
\eq{\begin{array}{@{}c@{}}\fwbox{50pt}{1}\\[-5pt]
\fwbox{50pt}{{\color{dim}0}}\fwbox{50pt}{{\color{dim}0}}\\[-5pt]
\fwbox{50pt}{{\color{dim}0}}\fwbox{50pt}{1}\fwbox{50pt}{{\color{dim}0}}\\[-5pt]
\fwbox{50pt}{{\color{dim}0}}\fwbox{50pt}{{\color{dim}0}}\fwbox{50pt}{{\color{dim}0}}\fwbox{50pt}{{\color{dim}0}}\\[-5pt]
\fwbox{50pt}{{\color{dim}0}}\fwbox{50pt}{{\color{dim}0}}\fwbox{50pt}{1}\fwbox{50pt}{{\color{dim}0}}\fwbox{50pt}{{\color{dim}0}}\\[-5pt]
\fwbox{50pt}{{\color{dim}0}}\fwbox{50pt}{{\color{dim}0}}\fwbox{50pt}{{\color{dim}0}}\fwbox{50pt}{{\color{dim}0}}\fwbox{50pt}{{\color{dim}0}}\fwbox{50pt}{{\color{dim}0}}\\[-5pt]
\fwbox{50pt}{1}\fwbox{50pt}{38711}\fwbox{50pt}{965644}\fwbox{50pt}{2473326}\fwbox{50pt}{965644}\fwbox{50pt}{38711}\fwbox{50pt}{1}\\[-5pt]
\fwbox{50pt}{{\color{dim}0}}\fwbox{50pt}{{\color{dim}0}}\fwbox{50pt}{{\color{dim}0}}\fwbox{50pt}{{\color{dim}0}}\fwbox{50pt}{{\color{dim}0}}\fwbox{50pt}{{\color{dim}0}}\\[-5pt]
\fwbox{50pt}{{\color{dim}0}}\fwbox{50pt}{{\color{dim}0}}\fwbox{50pt}{1}\fwbox{50pt}{{\color{dim}0}}\fwbox{50pt}{{\color{dim}0}}\\[-5pt]
\fwbox{50pt}{{\color{dim}0}}\fwbox{50pt}{{\color{dim}0}}\fwbox{50pt}{{\color{dim}0}}\fwbox{50pt}{{\color{dim}0}}\\[-5pt]
\fwbox{50pt}{{\color{dim}0}}\fwbox{50pt}{1}\fwbox{50pt}{{\color{dim}0}}\\[-5pt]
\fwbox{50pt}{{\color{dim}0}}\fwbox{50pt}{{\color{dim}0}}\\[-5pt]
\fwbox{40pt}{1}
\end{array}.}
\end{itemize}
The structure of these Hodge diamonds is very simple; a nontrivial cohomology only exists for degrees \((p, p)\) and \((p, k-p-1)\), corresponding to their middle column and row.
Interestingly, the form of these Hodge diamonds is compatible with hypersurfaces embedded in ordinary (unweighted) projective space (see appendix~\ref{sec:lefschetz} for a short discussion). It would be interesting to understand why this occurs, as we currently do not know how to embed our hypersurfaces in unweighted projective space.

To further characterize the Calabi-Yau manifold $X$ in $\wpok$, we compute its Euler characteristic $\chi(X)$.
The Euler characteristic is equal to the alternating sum of the dimensions of the de Rham cohomology groups, \mbox{$\chi(X) = \sum_{m} (-1)^m \sum_{p + q = m} h^{p,q}$}.
Following ref.~\cite{Hubsch:1992nu}, we can obtain a closed expression for it using an index theorem, see appendix \ref{sec:euler_characteristic} for details.
We find 
\begin{align}
  \chi(X) = \frac{1 - (1 - 2k)^k + 2 k^2}{2 k}\,.
\end{align}
The Euler characteristic of $X$ for low values of $k$ is given in table~\ref{tab:euler-characteristic-data}.

\begin{table}[t]
  \centering
  \begin{tabular}{c|c c c c c c c c c}
  $k$ & $2$ & $3$ & $4$ & $5$ & $6$ & $7$ & $8$ & $9$ \\
  \hline
  $\chi(X)$ & $0$ & $24$ & $-296$ & $5910$ & $-147624$ & $4482044$ & $-160180656$ & $6588215370$ \\
  \end{tabular}
  \caption{Euler characteristic $\chi(X)$ of $(k-1)$-dimensional Calabi-Yau hypersurfaces $X$ in \wpok\, for low values of $k$.}
  \label{tab:euler-characteristic-data}\vspace{-6pt}
\end{table}

We have seen above that a codimension-one Calabi-Yau hypersurface $X$ in $\wpok$ is defined by a polynomial $Q(x_1, \ldots, x_k, y)$ of the form given in eqn.~\eqref{eq:defining-eq-hypersurface-wpok}.
In the examples considered in the following sections, we will find polynomials of precisely this form with different coefficients $c_{\vec{\alpha},\beta}$, i.e.\ with different complex structure moduli.
The complex structure moduli of $X$ are in principle determined by integrating the holomorphic form of maximal degree along a basis of cycles on the manifold.
While in practice this is a difficult problem, we still give an account of how this form is constructed.

On $\wpok$, the canonical $k$-form $\Omega_k$ is given by
\begin{align}
  \label{eq:wpok-form}
  \Omega_k = k \; y \left( \bigwedge_{n = 1}^{k} d x_n \right) + \sum_{n = 1}^{k} (-1)^n x_n \; d y \wedge \left( \bigwedge_{m \neq n} d x_m \right)\,.
\end{align}
The Calabi-Yau hypersurface $X$ is defined as the zero-locus of the polynomial $Q(x_1, \ldots, x_k, y)$ in eqn.~\eqref{eq:defining-eq-hypersurface-wpok}.
The holomorphic form $\omega_{k-1}$ of (maximal) degree $k-1$ on $X$ is then given by
\begin{align}
  \label{eq:cy-form-from-embedding}
  \omega_{k-1} = \operatorname{Res} \frac{\Omega_k}{Q}\,.
\end{align}
The residue above is determined\footnote{Outside of the hypersurface \(X\), this residue is not uniquely defined since we could add to \(\operatorname{Res} \frac {\Omega_k} Q\) terms proportional to \(Q\).  However, when pulled back to \(X\), these terms vanish.} by the property that
\begin{equation}
  \frac{\Omega_k}{Q} = \left(\operatorname{Res} \frac{\Omega_k}{Q}\right) \wedge \frac {d Q} Q + \cdots\,,
\end{equation}
where the omitted terms are regular on the surface \(Q = 0\).

The hypersurfaces we encounter in the following sections turn out not to be smooth---i.e.\ there are non-trivial solutions to the system of polynomial equations $Q(x_1, \ldots, x_k, y) = \operatorname{d} Q(x_1, \ldots, x_k, y) = 0$.
Heuristically, the reason for this is that some of the monomials that would in principle be allowed for homogeneous polynomials in the coordinates of $\wpok$ are missing in $Q$.
Moreover, the coefficients depend on a limited number of kinematic variables, which is usually much smaller than the number of complex structure moduli.
In order to regularize the polynomials arising during integration, we can however consider a deformation of the complex structure, i.e.\ of the coefficients of the $c_{\vec{\alpha},\beta}$ in eqn.~\eqref{eq:defining-eq-hypersurface-wpok}.
Equivalently, we may say that we are considering the polynomials that we encounter in the following examples as special cases of a generic (smooth) polynomial $Q$ as defined in eqn.~\eqref{eq:defining-eq-hypersurface-wpok}. We provide more details on desingularization by complex structure deformation in appendix~\ref{sec:desing_cplx_structure}.

\newpage\section{Three-Loop Integrals Involving Calabi-Yaus in \texorpdfstring{$\mathbb{WP}^{1,\ldots,1,k}$}{WP(1,...,1,k)}}\label{sec:three-loop}%

Among the growing list of examples of Feynman integrals involving Calabi-Yau geometries are those with surprisingly few propagators---such as the so-called `banana' integrals or `tardigrades',
\vspace{-5pt}\eq{\fig{-22pt}{1}{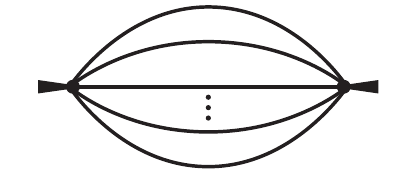}\qquad\text{and}\qquad\fig{-35pt}{1}{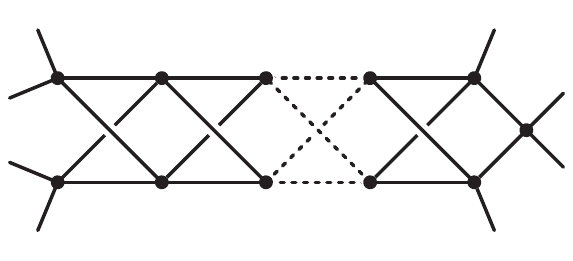}\,.\label{marginalCYs}\vspace{-5pt}}
These integrals are sub-topologies\footnote{We consider one Feynman integrand a sub-topology of another if the graph of the former's propagators is a quotient of the latter's by an (internal) edge contraction.} of almost all Feynman integrands at sufficiently high multiplicity, and it seems that any integral with a sub-topology involving a Calabi-Yau itself involves a Calabi-Yau. Thus, even for the special classes of scattering amplitudes that are expected to be polylogarithmic to all orders (see e.g. ref.~\cite{Arkani-Hamed:2014via}), it seems impossible that any local, Feynman-integrand-level representation can have this property term-by-term. Thus, it is essential that we learn to better understand these examples.

This sense of the ubiquity of Calabi-Yau geometries can be made more precise in the context of generalized unitarity, where it is possible to describe \emph{bases} of Feynman integrands subject to certain constraints. A basis large enough to represent all-multiplicity amplitudes in planar, maximally ($\mathcal{N}=4$) supersymmetric Yang-Mills (SYM) theory through three loops was described in ref.~\cite{Bourjaily:2017wjl}. Although $\mathcal{N}=4$ SYM theory in the planar limit is an unquestionably simple theory, this basis represents a necessary part of any larger basis needed to represent amplitudes in theories with ultraviolet behavior worse than $\mathcal{N}=4$ SYM theory (including the Standard Model). Thus, it is a natural place to start our understanding of the Calabi-Yau geometries relevant to general amplitudes.

At three loops, the basis of integrands needed for planar $\mathcal{N}=4$ SYM theory consists of the traintrack and wheel integrands shown in \mbox{figure~\ref{fig:planarCYs}}, and all irreducible integrands that contain one (or both) as sub-topologies and scale like either integrand (or better) in the ultraviolet. Thus, these two examples arise nearly ubiquitously (at large enough multiplicity) in three-loop amplitudes, motivating us in this section to study the Calabi-Yau geometry that arises in each. But first, let us describe the methods by which we may uncover these geometries.

\subsection{Identifying Calabi-Yau Geometries via Residues}\label{subsec:residues}%

Several infinite classes of Feynman diagrams have been shown to involve Calabi-Yau hypersurfaces in $\mathbb{WP}^{1,\ldots,1,k}$ using direct integration~\cite{Brown:2009ta,Bourjaily:2018yfy}. For instance, the two-dimensional banana graphs and four-dimensional tardigrades shown in eqn.~\eqref{marginalCYs} both fall into this category. In fact, these integral families both achieve the maximum possible degree of non-polylogarithmicity for marginal integrals. More precisely, the $L$-loop representative of each family saturates a bound on the possible `rigidity' of marginal integrals, where the rigidity of an integral is defined to be the dimension of the algebraic variety one must integrate over after a maximal number of polylogarithmic integrations have been carried out~\cite{Bourjaily:2018yfy}. The banana graphs have rigidity $L{-}1$, while the tardigrades have rigidity $2(L{-}1)$.\footnote{In the case of equal masses, the three-loop banana integral was recently expressed in terms of elliptic multiple polylogarithms \cite{Broedel:2019kmn}. While it involves a K3 surface, this K3 surface is related to the elliptic curve describing the two-loop sunrise graph in a way that drastically simplifies the problem \cite{Bloch:2014qca}. For general Calabi-Yau hypersurfaces, we would not expect this procedure to work, but it would be interesting to see to what extent it is possible. (See ref.~\cite{Besier:2019hqd} for some work in this direction.)}
The two-dimensional massive banana graphs are required, for example, in the calculation of the electron self-energy in QED \cite{Laporta:2008sx}, while the massless two-loop tardigrades enter the integrand basis for massless two-loop amplitudes using prescriptive unitarity \cite{Bourjaily:2013mma,Bourjaily:2015jna,Bourjaily:2017wjl,Bourjaily:2019iqr,integrandBases}.

In this work, we instead use sequences of residues to identify Calabi-Yau hypersurfaces in Feynman integrals, as done in ref.~\cite{Bourjaily:2018ycu}. In particular, we begin with representations of (here non-marginal) Feynman integrals at $L$ loops in terms of rational integrands involving only $2L$ integration variables (motivated by the conjectured bound of transcendental weight $2L$ at $L$ loops in four dimensions). We then examine the singular locus of these integrands by taking as many residues as we can.\footnote{If necessary, we perform changes of variables to rationalize square roots of quadratic polynomials along the lines of ref.~\cite{Besier:2018jen}.} This leads us to an expression of the form
\eq{
\frac{d \vec{x}}{\sqrt{P(\vec{x})}},
}
where $P(\vec{x})$ is a polynomial which is cubic or higher degree in the remaining variables without repeated roots. After projectivization, this polynomial defines a codimension-one hypersurface in $\wpok$ via eqn.~\eqref{eq:defining-eq-hypersurface-wpok-simplified}.

It is important to note that the above procedure mimics \emph{but is not equivalent to} the procedure of direct hyperlogarithmic integration. They are superficially similar in that direct integration partial-fractions rational integrands to isolate poles in the integration variable, while taking sequential residues also isolates poles. However, the partial-fractioning step of direct integration generates a term for each pole of the integrand, and preserves information about that pole in the form of the polylogarithmic function it constructs. If \textit{any} of these poles introduce a square root in the remaining variables, then this dependence will appear in the polylogarithmic integrand and direct integration may be obstructed. In contrast, by taking residues we may avoid this type of obstruction. As a result, the hypersurfaces we discuss in this section will not necessarily correspond to the degree of rigidity of the integrals involved; the integrals may be more `rigid' than the geometry we describe would suggest.

While our residue procedure does not necessarily uncover the maximally rigid geometry, it does uncover a geometry that is important and necessary to the understanding of these Feynman integrals. In particular, it is a geometry that should characterize the periods obtained by analytic continuation in the kinematics. To motivate this, recall that we can isolate any particular residue of the integrand with an integration contour tailored to that purpose. These closed integration contours represent potential ambiguities in the original Feynman integration contour, corresponding to the possibility to encircle additional branch cuts. Much as analytically continuing polylogarithmic functions around branch cuts results in factors of $2\pi i$, analytically continuing one of the integrals discussed in this work should give rise to integrals over the maximal residues we can perform---that is, integrals over the holomorphic forms of the Calabi-Yau manifolds we describe.

\subsection{Revisiting the Three-Loop Traintrack Integral}
\label{sec:three-loop-traintrack}

\begin{figure}[t]
\centering
\includegraphics{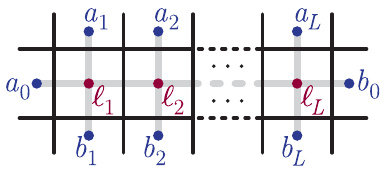} \qquad
\includegraphics{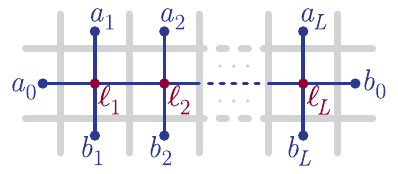}
\caption{The $L$-loop traintrack integral and its dual graph.}\vspace{-6pt}
\label{fig:traintrack}
\end{figure}

In ref.~\cite{Bourjaily:2018ycu}, some of the authors provided evidence that the $L$-loop traintrack integral, depicted in figure~\ref{fig:traintrack}, involves an integral over a Calabi-Yau $(L{-}1)$-fold. There, a manifestly dual-conformally invariant $2L$-fold representation was given for this integral:
\eq{\traintrack\!=\!\int\limits_0^{\infty}\!\!\proj{d^{L}\!\vec{\alpha}}d^L\!\vec{\beta}\frac{1}{\big(f_1\cdots f_L\big)g_L}\,,\label{dci_parametric_rep}}
where\footnote{Note that we have fixed a typo in $f_k$ from the published version of ref.~\cite{Bourjaily:2018ycu}.}
\vspace{52pt}
\eq{\fwbox{0pt}{\hspace{1pt}\,\begin{array}{@{}l@{}l@{}}~\\[-60pt]f_k\!\equivR&\u{a_0a_{k-1}}{a_kb_{k-1}}\hspace{-1pt}\u{a_{k-1}b_k}{b_{k-1}a_0}\hspace{-1pt}\u{a_kb_k}{b_{k-1}a_{k-1}}f_{k-1}\pl\alpha_0(\alpha_k\pl\beta_k)\pl\alpha_k\beta_k\,\\&\displaystyle\pl\!\sum_{j=1}^{k-1}\Big[\alpha_j\alpha_k\u{b_ja_0}{a_ja_k}%\\[-2pt]&\displaystyle\hspace{0pt}
\pl\alpha_j\beta_k\u{b_ja_0}{a_jb_k}\pl\alpha_k\beta_j\u{a_0a_j}{a_kb_j}\pl\beta_j\beta_k\u{a_0a_j}{b_kb_j}\hspace{-1pt}\Big]\!,\\[-2pt]g_L\!\equivR&\displaystyle\,\alpha_0\,\pl\!\sum_{j=1}^{L}\Big[\alpha_j\u{b_ja_0}{a_jb_0}\pl\beta_j\u{a_0a_j}{b_0b_j}\Big],\\[5pt]~\end{array}}\vspace{-14pt}\label{denominator_factors_defined}}
and $\u{{\color{hgreen}x}{\color{hred}y}}{{\color{hblue}z}{\color{hgreen}w}}$ denotes the cross-ratio
\eq{\u{{\color{hgreen}x}{\color{hred}y}}{{\color{hblue}z}{\color{hgreen}w}}\equivR\frac{\x{{\color{hgreen}x}}{{\color{hred}y}}\x{{\color{hblue}z}}{{\color{hgreen}w}}}{\x{{\color{hgreen}x}}{{\color{hblue}z}}\x{{\color{hred}y}}{{\color{hgreen}w}}}\,.\label{cross_ratio_notation_defined}}
The notation $\x{a}{b}\!\equivR\!(x_a\mi x_b)^2$ is intended to be suggestive of the embedding (or momentum-twistor) formalism.

We now specialize to three loops. Since each $f_k$ is linear in every integration variable, we can take residues in $\beta_1$, $\beta_2$, and $\beta_3$ on the locus of $f_1=f_2=f_3=0$. This leaves a single factor in the denominator, which is a rational function of $\alpha_0, \alpha_1, \alpha_2,$ and $\alpha_3$. Performing one final residue in $\alpha_3$, we obtain a square root of a polynomial with no repeated roots, $P_\mathfrak{T}(\alpha_0, \alpha_1,\alpha_2)$. This polynomial is degree six in $\alpha_0$ and $\alpha_1$ and degree four in $\alpha_2$ (the latter fact motivated the authors of ref.~\cite{Bourjaily:2018ycu} to put this polynomial into Weierstrass form with respect to $\alpha_2$, which will here prove unnecessary). Importantly, it can be checked that $P_\mathfrak{T}$ is a \textit{homogeneous} polynomial in $\alpha_0$, $\alpha_1$, and $\alpha_2$ of (overall) degree six. Therefore, writing this hypersurface as
\begin{equation}
Q(x_1,x_2,x_3,y)=y^2-P_\mathfrak{T}(x_1, x_2,x_3)=0,
\end{equation}
we identify it as a degree-six hypersurface in $\mathbb{WP}^{1,1,1,3}$. Generic surfaces of this type are well known to be K3 manifolds, which have Hodge diamond~\eqref{eq:K3_hodge_diamond} and Euler characteristic 24. We include the original three-loop integrand (from eqn.~\eqref{dci_parametric_rep}) in \textsc{Mathematica} format in the ancillary file \texttt{integrands\_and\_varieties.m}.

\subsection{The Three-Loop Wheel Integral}\label{sec:three-loop-wheel}

\begin{figure}[b]
\centering
\includegraphics{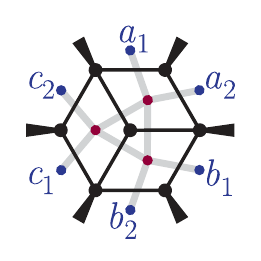} \qquad
\includegraphics{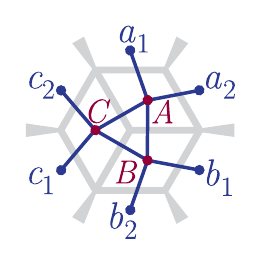}
\caption{The three-loop wheel integral and its dual graph.}
\label{fig:three_loop_wheel}
\end{figure}

The three-loop scalar wheel integral is drawn in momentum space and dual-momentum space in figure \ref{fig:three_loop_wheel}. Using the notation presented in the previous subsection, it is given by
\eq{\wheelint^{(3)}\hspace{5pt}\equivR\int\!\!\!\frac{\fwboxR{0pt}{d^4\!x_{\!{\color{hred}A}}d^4\!x_{\!{\color{hred}B}}d^4\!x_{\hspace{-0.5pt}{\color{hred}C}}\,\hspace{6pt}}\x{a_1}{a_2}\hspace{-0.75pt}\x{b_1}{b_2}\hspace{-0.75pt}\x{c_1}{c_2}}{\x{{\color{hred}A}}{{\color{hred}C}}\x{{\color{hred}A}}{a_1}\hspace{-0.75pt}\x{{\color{hred}A}}{a_2}\hspace{-0.75pt}\x{{\color{hred}A}}{{\color{hred}B}}\hspace{-0.75pt}\x{{\color{hred}B}}{b_1}\hspace{-0.75pt}\x{{\color{hred}B}}{b_2}\hspace{-0.75pt}\x{{\color{hred}B}}{{\color{hred}C}}\hspace{-0.75pt}\x{{\color{hred}C}}{c_1}\hspace{-0.75pt}\x{{\color{hred}C}}{c_2}\hspace{-0.75pt}}\,,\label{wheel_spacetime_integral}}
where we have included a numerator that renders it dual-conformally invariant. In appendix \ref{appendix:derivation_of_the_wheel}, we derive an equivalent six-fold integral representation of this integral, following the strategy of refs.~\cite{Bourjaily:2019jrk,Bourjaily:2018ycu,Bourjaily:2018aeq}. We quote the result here for convenience:
\eq{\wheelint^{(3)}=\int\limits_0^\infty\!\!\!d^2\!\vec{\alpha}\,d^2\!\vec{\beta}\,d^2\!\vec{\gamma}\,\,\frac{n_0}{f_1\,f_2\,f_3},
\label{eq:wheelint}}
where
\vspace{-46pt}\eq{\begin{split}~\\[20pt]
n_0\equivR&\hspace{-4pt}\phantom{+}\xrv_1(\xru_1\xru_2\xru_3\xrv_1\xrv_2\xrv_3)\,,\\
f_1\equivR&\hspace{-4pt}\phantom{+}\alpha_1+\alpha_2+\alpha_1\alpha_2\,,\\
f_2\equivR&\hspace{-4pt}\phantom{+}\alpha_1(1+\alpha_2+\beta_1+\beta_2+\gamma_2)+\alpha_2(1+ \xru_1 \xrw_2(\xrw_3\beta_1+\beta_2)+\gamma_2)\\
&\hspace{-4pt}\!+\!\beta_1\xrv_1(1+ \xru_1\xru_3\xrv_2\xrw_2\beta_2+\gamma_2)+ \xru_2\xrv_1(\xru_1\xrv_3\gamma_1+\beta_2(1+\gamma_2))\,,\\
f_3\equivR&\fwboxL{0pt}{\hspace{-8pt}\phantom{+}(1\!+\!\alpha_2\!+\!\beta_1\!+\!\beta_2\!+\!\gamma_2)\Big[\alpha_1\hspace{-2pt}\Big(\hspace{-2pt}\gamma_1\!\!+\!\beta_2(1\!+\!\alpha_2\!+\! \xru_3\xrv_1\xrv_2\beta_1\!+\!\gamma_2)\!+\hspace{-1pt}\xrw_3\beta_1\hspace{-1pt}(1\!+\!\alpha_2\!+\!\hspace{-1pt}\gamma_2)\hspace{-2pt}\Big)}\\
&\hspace{-4pt}\!+\!(1\!+\!\hspace{-1pt}\gamma_2)\big(\hspace{-2pt}\xrw_3\alpha_2\beta_1\!+\!(\alpha_2\!+\!\xru_3 \xrv_1\xrv_2\beta_1)\beta_2\hspace{-2pt}\big)\hspace{-2pt}\Big]\!+\!\gamma_1\Big[\alpha_2(1\!+\!\xru_1(\xrw_3\beta_1\!+\!\beta_2)\!\!+\!\!\gamma_2)\\
&\hspace{-4pt}\!+\!\xru_3 \xrv_1(\xru_2 \xrw_1\beta_2(1\!+\!\gamma_2)\!+\!\beta_1(1\!+\!\xru_1 \xrv_2\beta_2\!+\!\gamma_2))\Big]\,, ~\label{feynman_param_int_denominators}
%\\[-0pt]
\end{split}
}
and where we have used the following basis of dual-conformal invariant cross-ratios:
\eq{\hspace{-70pt}\begin{array}{@{}r@{}l@{$\hspace{5pt}$}r@{}l@{$\hspace{5pt}$}r@{}l@{}}\xru_1\!\equivR&\!\u{c_1a_1}{a_2b_2}\,,&\xru_2\!\equivR&\!\u{a_1b_1}{b_2c_2}\,,&\xru_3\!\equivR&\!\u{b_1c_1}{c_2a_2}\,,\\
\xrv_1\!\equivR&\!\u{a_1a_2}{b_1c_2}\,,&\xrv_2\!\equivR&\!\u{b_1b_2}{c_1a_2}\,,&\xrv_3\!\equivR&\!\u{c_1c_2}{a_1b_2}\,,\\
\xrw_1\!\equivR&\!\u{b_2c_1}{c_2b_1}\,,&\xrw_2\!\equivR&\!\u{c_2a_1}{a_2c_1}\,,&\xrw_3\!\equivR&\!\u{a_2b_1}{b_2a_1}\,.\end{array}\hspace{-50pt}\label{general_cross_ratios_defined}}
Note that the dihedral symmetry of $\wheelint^{(3)}$ acts quite naturally on these variables. Specifically, under the dihedral group that leaves the graph in figure \ref{fig:three_loop_wheel} invariant, the $u_i$'s, $v_i$'s and $w_i$'s each form a three-orbit. We include this integrand in \textsc{Mathematica} format in the ancillary file \texttt{integrands\_and\_varieties.m}.

To analyze the geometry of $\wheelint^{(3)}$~\eqref{eq:wheelint}, we first take three residues on the locus \(f_i = 0\) by eliminating the variables $\alpha_1$, $\beta_2$, and $\gamma_1$. We thereby obtain a three-form
\begin{equation}
  \frac {d \alpha_2 d \beta_1 d \gamma_2}{\sqrt{P_\wheelint(\alpha_2, \beta_1, \gamma_2)}},
\end{equation}
where \(P_\wheelint\) is a \emph{non-homogeneous} polynomial.  However, assigning \(\alpha_2\), \(\beta_1\), and \(\gamma_2\) all weight one, we can homogenize \(P_\wheelint(\alpha_2, \beta_1, \gamma_2)\) by adding a fourth (auxiliary) weight-one coordinate \(x_4\). The resulting homogeneous polynomial can be chosen to have overall degree eight, and we denote it \(P^8_\wheelint(\alpha_2, \beta_1, \gamma_2, x_4)\).  As it is rather long, we do not present this polynomial in the text, but we provide it in the ancillary file \texttt{integrands\_and\_varieties.m}. Finally, introducing a weight-four variable $y$ with \(y^2 = P^8_\wheelint(x_1, x_2, x_3, x_4)\), we obtain a three-form which can be expressed as
\begin{equation}
  \omega_3=\frac {x_4 d x_1 d x_2 d x_3} y
\end{equation}
in the patch where \(x_4\) is a non-vanishing constant.

Up to a numerical factor, the three-form $\omega_3$ can be obtained from eqn.~\eqref{eq:cy-form-from-embedding} by taking the residue of
\begin{equation}
  \frac {\Omega_{4}}{y^2 - P^8_\wheelint(x_1, x_2, x_3, x_4)}
\end{equation}
at the locus defined by the vanishing of the denominator, where \(\Omega_{4}\) is the canonical four-form on \(\mathbb{WP}^{1, 1, 1, 1, 4}\) given in eqn.~\eqref{eq:wpok-form},
\begin{align}
\Omega_4 &= 4 y d x_1 d x_2 d x_3 d x_4 +
  d y \Big(-x_1 d x_2 d x_3 d x_4
      +x_2 d x_1 d x_3 d x_4 \\
      &\hspace{5cm}
       -x_3 d x_1 d x_2 d x_4
       +x_4 d x_1 d x_2 d x_3 \Big). \nonumber
\end{align}
It follows that \(Q(x_1, x_2, x_3, x_4,y)=y^2 - P^8_\wheelint(x_1, x_2, x_3, x_4) = 0\) defines a Calabi-Yau threefold in \(\mathbb{WP}^{1, 1, 1, 1, 4}\).  The polynomial \(P^8_\wheelint\) has \(\binom{8 + 3}{3} = 165\) coefficients, which can be parametrized by \(\binom{8 + 3}{3}- 16= 149\) complex structure moduli, but in our case they depend only on the nine cross-ratios in eqn.~\eqref{general_cross_ratios_defined}.  Hence, by varying these cross-ratios, we only explore a small part of the complex structure moduli space of our Calabi-Yau threefold.

\subsubsection*{Interesting Kinematic Limits}

We start by considering the limit in which the legs at the rungs of the wheel become massless. This corresponds to the condition that the dual coordinates on either side of these legs become light-like separated, namely $\x{a_2}{b_1} \to 0$, $\x{b_2}{c_1} \to 0$, and $\x{c_2}{a_1} \to 0$.
In the variables~\eqref{general_cross_ratios_defined}, this sets all three parameters $\xrw_i= 0$:
\eq{\big\{\x{a_2}{b_1}\to0,\;\x{b_2}{c_1}\to0,\;\x{c_2}{a_1}\to0\big\}\quad\bigger{\Leftrightarrow}\quad\big\{\xrw_1\to0,\;\xrw_2\to0,\;\xrw_3\to0\,\big\}\label{massless_middle_limit_defn}}
\vspace{-14pt}\eq{\fig{-24.65pt}{1}{figures/wheel_integral}\bigger{\underset{\text{(\ref{massless_middle_limit_defn})}}{\Longrightarrow}}\fig{-35.25pt}{1}{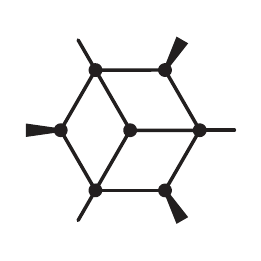}\bigger{\Leftrightarrow}\fig{-35.25pt}{1}{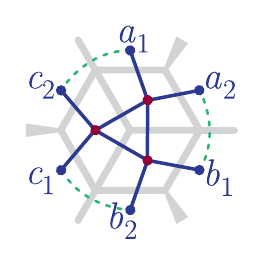}\label{massless_middle_limit}\vspace{-0pt}}
(Notice that we denote light-like separated points in the dual graph by dashed green lines.) It can be checked that the resulting integral is still a Calabi-Yau hypersurface in \(\mathbb{WP}^{1, 1, 1, 1, 4}\).

To see this Calabi-Yau threefold factorize into simpler geometries, we now consider the limit in which one of these massless legs becomes soft. It can easily be checked that identifying $a_2 = b_1$ sets $\xru_3=\xrv_1=\xrv_2=1$:
\eq{\big\{\xru_3\to1,\;\xrv_1\to1,\;\xrv_2\to1,\;\xrw_i\to0\big\}\label{empty_corner_limit_defn}}
\vspace{-14pt}\eq{\fig{-35.25pt}{1}{figures/wheel_integral_degn1}\bigger{\underset{\text{(\ref{empty_corner_limit_defn})}}{\Longrightarrow}}\fig{-35.25pt}{1}{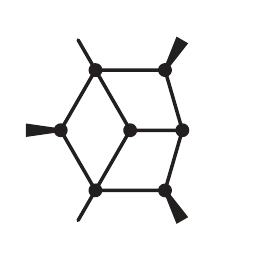}\bigger{\Leftrightarrow}\fig{-35.25pt}{1}{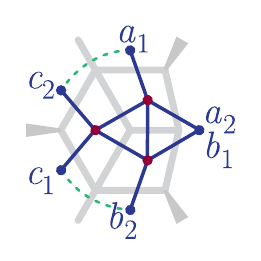}\label{empty_corner_limit}\vspace{-0pt}}
In this limit, $P(\alpha_2, \beta_1, \gamma_2)$ factorizes, and one of its factors is a perfect square. This allows us to take an additional residue.
Continuing on in this fashion, we find we can take residues in all six integration variables; so from the residue analysis, there is no irreducible geometry. However, direct integration is obstructed after just a single integration; as we emphasized in section \ref{subsec:residues}, these functions may appear to be more rigid under direct integration than their residue analysis would suggest.

It turns out this obstruction can be avoided by additionally setting $\xrv_3=1$ (in this case our choice is purely pragmatic, and not particularly motivated by physics). On this kinematic slice, the integral evaluates to
\eq{\begin{split}
\hspace{-15pt}\frac{u_1 u_2}{u_1-u_2}\Big\{&\phantom{-\,}\mpl{0,1,1,0,1,0}{1}+\mpl{0,1,0,1,0,0}{1}+\mpl{0,0,1,0,0,0}{1}+\mpl{0,0,0,1,1,0}{1}-\mpl{0,1,1,0,0,0}{1}-\mpl{0,1,0,1,1,0}{1}\hspace{-20pt}\\[-4pt]
&\!-\!\mpl{0,0,1,0,1,0}{1}-\mpl{0,0,0,1,0,0}{1}+\mpl{0}{2}\Big(\mpl{1,1,0,0,0}{1}+\mpl{1,0,1,1,0}{1}+\mpl{0,1,0,1,0}{1}+\mpl{0,0,1,0,0}{1}\hspace{-20pt}\\[-2pt]
&\!-\!\mpl{1,1,0,1,0}{1}-\mpl{1,0,1,0,0}{1}-\mpl{0,1,0,0,0}{1}-\mpl{0,0,1,1,0}{1}\Big)+\Big(\mpl{1,0}{2}-\mpl{0,0}{2}\Big)\Big(\mpl{1,0,1,0}{1}\\[-2pt]
&\!-\!\mpl{1,0,0,0}{1}-\mpl{0,1,1,0}{1}+\mpl{0,1,0,0}{1}\Big)-\Big(\mpl{1,1,0}{2}-\mpl{1,0,0}{2}\Big)\Big(\mpl{0,1,0}{1}-\mpl{0,0,0}{1}\Big)\\[-2pt]
&\!+\!\zeta_2\Big[\mpl{1,0,0,0}{1}+\mpl{0,1,0,1}{1}+\mpl{0,0,1,0}{1}-\mpl{1,0,1,0}{1}-\mpl{0,1,0,0}{1}-\mpl{0,0,0,1}{1}\\[-2pt]
&\hspace{24pt}+\mpl{1}{2}\Big(\mpl{0,1,0}{1}-\mpl{0,0,0}{1}\Big)-\mpl{0}{2}\Big(\mpl{1,0,1}{1}-\mpl{1,0,0}{1}+\mpl{0,1,0}{1}-\mpl{0,0,1}{1}\Big)\\[-2pt]
&\hspace{24pt}+\mpl{1,0}{2}\Big(\mpl{0,1}{1}-\mpl{0,0}{1}\Big)-\mpl{0,0,}{2}\mpl{0,1}{1}\Big]+2\zeta_3\Big[\mpl{1,0,0}{1}-\mpl{1,1,0}{1}-\mpl{0,1,1}{1}+\mpl{0,0,1}{1}\hspace{-28pt}\\[-2pt]
&\!+\!\mpl{1}{2}\Big(\mpl{1,0}{1}-\mpl{0,0}{1}\Big)+\mpl{0}{2}\Big(\mpl{1,1}{1}-\mpl{0,1}{1}\Big)\Big]-\frac{7}{5}\zeta_2^2\Big(\mpl{1,0}{1}-\mpl{0,1}{1}+\mpl{0}{2}\mpl{1}{1}\Big)\hspace{-28pt}\\[-4pt]
&\!+\!4\,\zeta_2\,\zeta_3\mpl{1}{1}\Big\}+\big(u_1\leftrightarrow u_2\big),\end{split}\label{five_point_wheel_slice}}
using the shorthand $G_{\!\vec{w}}^{\,z}\equivR G(\{\vec{w}\},z)$. We also include this expression in the ancillary file \texttt{integrands\_and\_varieties.m}.

Further simplifications may be achieved by taking a second of the massless legs to be soft. Identifying $a_1 = c_2$ after taking the limit~\eqref{empty_corner_limit_defn} additionally sets $\xru_2=\xrv_3=1$, making this integral the $\xru_2 \to 1$ limit of expression~\eqref{five_point_wheel_slice}:
\vspace{-14pt}\eq{\fig{-35.25pt}{1}{figures/wheel_integral_degn2}\bigger{\underset{\substack{u_2\to1\\v_3\to1}}{\Longrightarrow}}\fig{-35.25pt}{1}{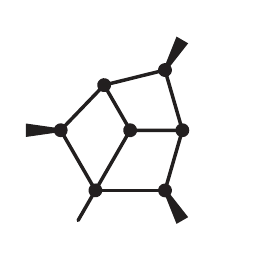}\bigger{\Leftrightarrow}\fig{-35.25pt}{1}{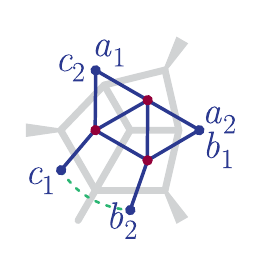}\label{double_empty_corner_limit}\vspace{-14pt}}
where
\eq{\begin{split}~\\[-24pt]
\hspace{-20pt}\fig{-35.25pt}{1}{figures/wheel_integral_degn3}\hspace{-16.5pt}=\frac{\xru_1}{1-\xru_1}\Big[&\phantom{-\,}\mpl{0, 1, 1, 0, 0, 0}{1}+\mpl{0, 1, 0, 1, 1, 0}{1}+\mpl{0, 0, 1, 0, 1, 0}{1}+\mpl{0, 0, 0, 1, 0, 0}{1}\\[-28pt]
&\!-\!\mpl{0, 1, 1, 0, 1, 0}{1}-\mpl{0, 1, 0, 1, 0, 0}{1}-\mpl{0, 0, 1, 0, 0, 0}{1}-\mpl{0, 0, 0, 1, 1, 0}{1}\\
&\!+\!\zeta_2\Big(\mpl{0, 1, 1, 0}{1}-\mpl{0, 1, 0, 1}{1}-\mpl{0, 0, 1,0}{1}+\mpl{0, 0, 0, 1}{1}\Big)\\
&\!+\!2\zeta_3\Big(\mpl{0, 1, 1}{1}+\mpl{0, 1, 0}{1}-\mpl{0,0,1}{1}-\mpl{0,0,0}{1}\Big)-6\zeta_4\Big(\mpl{0,1}{1}-\mpl{0,0}{1}\Big)\hspace{-20pt}\\
&\!-\!2(5 \zeta_5+ \zeta_2 \zeta_3 )\mpl{0}{1}+4( \zeta_2^3 -\zeta_3^2 ) + 3 \zeta_6\Big]\;.
\\[-5pt]\end{split}\label{wheel_on_line}
}
We also include this expression in the ancillary file \texttt{integrands\_and\_varieties.m}.

The last massless leg is removed by setting the final cross-ratio $u_1 = 1$:
\vspace{-14pt}\eq{\fig{-35.25pt}{1}{figures/wheel_integral_degn3}\hspace{-8pt}\bigger{\underset{u_1\to1}{\Longrightarrow}}\fig{-35.25pt}{1}{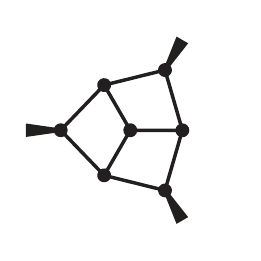}\hspace{-10pt}\bigger{\Leftrightarrow}\fig{-35.25pt}{1}{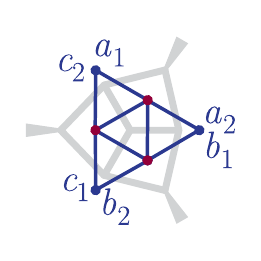}\label{hexagon_limit}\vspace{-10pt}}
In this limit, the integral evaluates to
\eq{
\fig{-35.25pt}{1}{figures/wheel_integral_degn4}\hspace{-20pt}=20\zeta_5.
}
This might na\"ively be surprising, as one expects the three-loop wheel to have transcendental weight six. However, one can observe that the rational prefactor diverges in the $u_1\rightarrow 1$ limit of expression~\eqref{double_empty_corner_limit}; in order to take this limit one should therefore expand the polylogarithmic part of this function in a power series, which leads to a drop in weight~\cite{Broadhurst:1995km,Schnetz:2008mp,Drummond:2012bg,Brown:2012ia,Bogner:2014mha,Panzer:2015ida,Golz:2015rea,Panzer:2016snt,Broadhurst:2016kho}.

\subsubsection*{A Three-Parameter Toy Model}

The three-loop wheel integral allows for a three-parameter toy model similar to that of the elliptic double-box \cite{Bourjaily:2017bsb}.
This toy model is defined by taking all six dual-momentum points defining the three-loop wheel integral to be light-like separated in sequence. That is, we take
\eq{\x{a_1}{b_2}=\x{b_2}{c_1}=\x{c_1}{a_2}=\x{a_2}{b_1}=\x{b_1}{c_2}=\x{c_2}{a_1}=0\,.\label{toy_model_limit_defn}}
\vspace{-10pt}\eq{\fig{-35.25pt}{1}{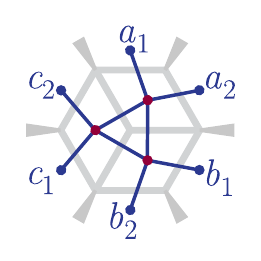}\bigger{\underset{\text{(\ref{toy_model_limit_defn})}}{\Longrightarrow}}\fig{-35.25pt}{1}{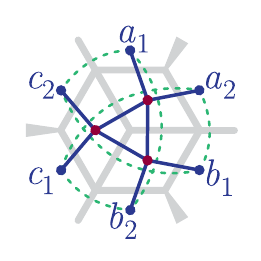}\label{toy_model_limit}\vspace{-0pt}}
In this limit, some of the rescalings of the Feynman parameters in our derivation become singular.\footnote{%
Concretely, the rescalings of the Feynman parameters $\beta_2$ taken in eqn.~\eqref{wheel_beta_change} and those for $\gamma_i$ in eqn.~\eqref{wheel_gamma_change} are singular in the limit \eqref{toy_model_limit_defn}. However, this observation clearly signals how these problems can be remedied: to access this limit smoothly from our previous expression \eqref{wheel_ult}, we merely need to rescale
\eq{\hspace{16pt}\begin{array}{@{}l@{}l@{$\hspace{20pt}$}l@{}l@{}}\beta_2&\displaystyle\mapsto\beta_2\frac{\x{a_1}{b_2}\x{a_2}{c_2}}{\x{a_1}{a_2}\x{b_2}{c_2}},
&\gamma_1&\displaystyle\mapsto\gamma_1\frac{\x{a_2}{c_1}\x{b_2}{c_2}}{\x{a_2}{b_2}\x{c_1}{c_2}},\\
\gamma_2&\displaystyle\mapsto\gamma_2\frac{\x{a_1}{a_2}\x{b_2}{c_2}}{\x{a_1}{b_2}\x{a_2}{c_2}},
&\gamma_3&\displaystyle\mapsto\!\hspace{-1pt}1\!\!\times\!\!\hspace{-1pt}\frac{\x{a_1}{a_2}\x{b_2}{c_2}}{\x{a_1}{b_2}\x{a_2}{c_2}}\,,\end{array}\label{to_wheel_toy_rescaling}}
take into account the relevant Jacobians, and collect terms. After this has been done, the limit \eqref{toy_model_limit_defn} can be taken smoothly.} The cross-ratios chosen in eqn.~\eqref{general_cross_ratios_defined} also become problematic; individually we have $v_i, w_i \to 0$, $u_i\to\infty$, while the ratios
\eq{t_1\equivR\frac{1}{\xru_1\xrv_2\xrv_3}=\u{b_1c_1}{b_2c_2},\;\;t_2\equivR\frac{1}{\xru_2\xrv_3\xrv_1}=\u{a_1c_1}{a_2c_2},\;\;t_3\equivR\frac{1}{\xru_3\xrv_1\xrv_2}=\u{a_1b_1}{a_2b_2}\label{toy_model_cross_ratios}}
remain finite. Accounting for both of these issues, we find the six-fold integral representation becomes
\eq{\wheelint^{(3)}\hspace{2pt}\underset{\substack{\text{(\ref{to_wheel_toy_rescaling})}\\\text{(\ref{toy_model_limit_defn})}}}{\longmapsto}\wheelint^{\text{toy}}\equivR\int\limits_0^\infty\!\!\!d^2\!\vec{\alpha}\,\,d^2\!\vec{\beta}\,\,d^2\!\vec{\gamma}\,\,\frac{1}{g_1\,g_2\,g_3}\,,\label{toy_wheel_int_ult}\vspace{-10pt}}
where
\eq{\begin{split}g_1\equivR&\phantom{\pl}\alpha_1\pl\alpha_2\pl\alpha_1\alpha_2\,,\\
g_2\equivR&\phantom{\pl}\alpha_1(1\pl\alpha_2\pl\beta_1\pl\gamma_2)\pl(\alpha_2\pl\beta_2)(1\pl\gamma_2)\pl\gamma_1\,,\\
g_3\equivR&\phantom{\pl}\alpha_1\beta_2(1\pl\alpha_2\pl\beta_1\pl\gamma_2)(\beta_1\pl {\color{hblue}\toyxr_3}(1\pl\alpha_2\pl\gamma_2))\pl\gamma_1\big[{\color{hblue}\toyxr_1}\beta_1(1\pl\gamma_2)\pl {\color{hblue}\toyxr_2}({\color{hblue}\toyxr_3}\alpha_2\pl\beta_1)\beta_2\big]\\
&\pl\beta_2(1\pl\alpha_2\pl\beta_1\pl\gamma_2)({\color{hblue}\toyxr_3}\alpha_2\pl\beta_1)(1\pl\gamma_2)\,,
\end{split}\label{toy_denominator_factors_defined_appendix}}
in terms of the cross-ratios~\eqref{toy_model_cross_ratios}.

As before, we can take residues in $\alpha_1$, $\beta_2$, and $\gamma_1$, obtaining a non-homogeneous curve $P^{\text{toy}}_\wheelint(\alpha_2,\beta_1,\gamma_2) = P^{\text{toy}}_\wheelint(x_1,x_2,x_3)$ that we can then homogenize with an auxiliary variable $x_4$. The resulting degree-eight polynomial is
\begin{align}
\hspace{-50pt}P^{8,\text{toy}}_\wheelint=\hspace{0pt}&\hspace{-0pt}\fwboxL{300pt}{\Big[x_2(x_1^2x_3-x_1x_2x_4){\color{hblue}t_2}+x_1^2\big(x_1 x_3({\color{hblue}t_2}-1)-(x_2+x_3)x_3-(x_2{\color{hblue}t_2}+x_3)x_4\big){\color{hblue}t_3}}\nonumber\\
&\phantom{\Big[}\,-x_2(x_1+x_2+x_3+x_4)(x_1x_3+(x_3+x_4)x_4)\Big]^2-2{\color{hblue}t_1}x_2(x_1+x_4)(x_3+x_4)^2\hspace{-30pt}\nonumber\\
&\times\Big[(x_1+x_2+x_3+x_4)\big(x_1^2\,x_3\,{\color{hblue}t_3}+x_1x_2x_3+x_2x_4(x_3+x_4)\big)\label{three_loop_wheel_toy_curve}\\
&\phantom{\Big[}\,\;\;+{\color{hblue}t_2}x_1(x_1{\color{hblue}t_3}+x_2)(x_1x_3-x_2x_4)\Big]+{\color{hblue}t_1^2}x_2^2(x_1+x_4)^2(x_3+x_4)^4\,.\nonumber
\end{align}
We include both the toy model integrand and the above hypersurface in \textsc{Mathematica} format in the ancillary file \texttt{integrands\_and\_varieties.m}.

We pause here to highlight that it is possible to see this polynomial factorize into simpler polynomials in simple kinematic limits. Despite its presentation, the toy model's \emph{geometry} must be invariant under permutations of $t_1$, $t_2$, and $t_3$; thus, we may consider taking limits in any variable. However, these limits can na\"ively look different; for instance, if we set $t_1 \to 0$ or $t_2 \to 0$, the polynomial becomes a perfect square of a polynomial with overall degree four, while in the limit $t_3 \to 0$ it factorizes into $x_2^2$ times a polynomial of overall degree six.
By symmetry, the irreducible geometry in each of these limits must be the same. In the first case (taking the limit in either $t_1$ or $t_2$), the resulting (squared) polynomial has degree three in $x_1$ and $x_4$, and degree two in $x_2$ and $x_3$. This lets us perform an additional residue in either $x_2$ or $x_3$, by which we obtain a square root of a polynomial of overall degree six. In the second case (taking the limit in $t_3$) we instead take a residue at $x_2=0$, after which the remaining polynomial has overall degree six. Both of the resulting polynomials define a K3, although it is not easy to see that they describe the same geometry (i.e.~that they correspond to different parametrizations of the same hypersurface).

If we take an additional cross-ratio to zero, the curve degenerates again. It becomes a square of a polynomial that is cubic in one variable and quadratic in the remaining two. This allows an additional residue in one of the quadratic variables, giving rise to a square root of a quadratic polynomial in the remaining two variables. Such square roots are rationalizable under a change of variables, so the integral should be polylogarithmic in this limit.

\vspace{-0pt}%
%\newpage
\section{Open Problems at Four Loops and Beyond}\label{sec:open-problems}%
\vspace{-0pt}

Having shown that the three-loop traintrack and wheel both involve Calabi-Yau hypersurfaces that can be embedded in $\mathbb{WP}^{1,\ldots,1,k}$, it is natural to ask whether their four-loop counterparts also involve such hypersurfaces.

\subsection*{The Four-Loop Traintrack}

Equations~\eqref{dci_parametric_rep} and \eqref{denominator_factors_defined} provide an eight-fold integral representation of the four-loop traintrack integral (which we again provide in \textsc{Mathematica} format in the ancillary file \texttt{integrands\_and\_varieties.m}). We can analyze the residues of this integral in the same way as was done for the three-loop integrals in the last section, to see if it contains a Calabi-Yau hypersurface in $\wpok$. Here we can take four residues, in $\beta_1, \beta_2, \beta_3,$ and $\beta_4$, on the locus $f_1=f_2=f_3=f_4=0$, then one final residue in $\alpha_4$ on $g_4=0$ to obtain a square root of a polynomial $P^{(4)}_\mathfrak{T}(\alpha_0,\alpha_1,\alpha_2,\alpha_3)$ with no repeated roots. This polynomial is homogeneous, but has overall degree \textit{ten}. It is degree ten in $\alpha_0$ and $\alpha_1$, degree six in $\alpha_2$, and degree four in $\alpha_3$. Taking other sequences of residues also result in degree ten, twelve, or sixteen polynomials.

As $P^{(4)}_\mathfrak{T}(\alpha_0,\alpha_1,\alpha_2,\alpha_3)$ has degree ten, $y^2-P^{(4)}_\mathfrak{T}(\alpha_0,\alpha_1,\alpha_2,\alpha_3)=0$ cannot be embedded in \(\mathbb{WP}^{1, 1, 1, 1, 4}\). It can be embedded in weighted projective space \(\mathbb{WP}^{1, 1, 1, 1, 5}\), but this does not satisfy the Calabi-Yau condition. We currently know of no way to embed this variety in a weighted projective space so that it satisfies the Calabi-Yau condition.

\subsection*{The Four-Loop Wheel}

The four-loop scalar wheel (or `window') integral, $\wheelint^{(4)}$, may be drawn in momentum space and dual-momentum space as
\vspace{-6pt}\begin{align}&\fwbox{200pt}{\hspace{14pt}\fwboxR{0pt}{\wheelint^{(4)}\equivR\hspace{-5pt}}\fig{-35.25pt}{1}{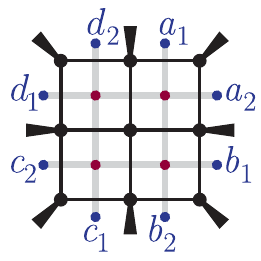}\bigger{\Leftrightarrow}\fig{-35.25pt}{1}{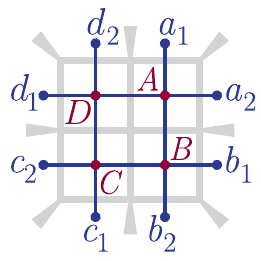}}\label{window_figure_intro}\\
&\fwboxL{200pt}{\hspace{-84pt}=\!\!\int\!\!\!\frac{\fwboxR{0pt}{d^4\!x_{\!{\color{hred}A}}d^4\!x_{\!{\color{hred}B}}d^4\!x_{\hspace{-0.5pt}{\color{hred}C}}d^4\!x_{\hspace{-0.5pt}{\color{hred}D}}\,\hspace{10pt}}\x{a_1}{a_2}\hspace{-0.75pt}\x{b_1}{b_2}\hspace{-0.75pt}\x{c_1}{c_2}\x{d_1}{d_2}}{\x{{\color{hred}D}}{{\color{hred}A}}\x{{\color{hred}A}}{a_1}\hspace{-0.75pt}\x{{\color{hred}A}}{a_2}\hspace{-0.75pt}\x{{\color{hred}A}}{{\color{hred}B}}\hspace{-0.75pt}\x{{\color{hred}B}}{b_1}\hspace{-0.75pt}\x{{\color{hred}B}}{b_2}\hspace{-0.75pt}\x{{\color{hred}B}}{{\color{hred}C}}\hspace{-0.75pt}\x{{\color{hred}C}}{c_1}\hspace{-0.75pt}\x{{\color{hred}C}}{c_2}\hspace{-0.75pt}\x{{\color{hred}C}}{{\color{hred}D}}\hspace{-0.75pt}\x{{\color{hred}D}}{d_1}\hspace{-0.75pt}\x{{\color{hred}D}}{d_2}\hspace{-0.75pt}}\,,}\label{window_spacetime_integral}
\vspace{-6pt}\end{align}
where in the last line we have written the integral explicitly in dual-momentum space. We derive a manifestly dual-conformally invariant integral representation of the four-loop wheel integral in general kinematics in appendix \ref{appendix:derivation_of_the_wheel_four}, finding
\eq{\wheelint^{(4)}=\int\limits_0^\infty\!\!\!d^2\!\vec{\alpha}\,\,d^2\!\vec{\beta}\,\,d^2\!\vec{\gamma}\,\,d^3\!\vec{\delta}\frac{n_0}{f_1\,f_2\,f_3}\left(\frac{n_1}{f_2}+\frac{n_2}{f_3}\right)\,.
\label{eq:windowint}}
The expressions for $n_0, n_1, n_2, f_1, f_2,$ and $f_3$ are lengthy, but are given in \textsc{Mathematica} format in the ancillary file \texttt{integrands\_and\_varieties.m}.

Unfortunately, this expression is a \emph{nine}-fold integral, while considerations of transcendental weight suggest that it should be possible to write down an eight-fold representation. This has direct consequences for the validity of our residue analysis. In particular, it means that we cannot directly associate the number of remaining integration parameters after taking a maximum number of residues with the dimension of an irreducible geometry. With this proviso in mind, we can take residues in $\alpha_1, \delta_1, \beta_1,$ and  $\alpha_2$, leaving a quartic with no repeated roots in five (non-projective) variables. This means that the geometry is at most a fivefold hypersurface, but could be of lower dimension. Without an eight-fold integral representation, we cannot distinguish these possibilities.

This integral has several limits with applications to integrable theories, which would make it particularly interesting to compute. We discuss these limits (some of which are polylogarithmic), as well as a nine-parameter toy model similar to the three-loop toy model~\eqref{toy_model_limit_defn}, in appendix~\ref{subsec:window_limits}.

\subsection*{Further Directions}

There are many open questions regarding the types of varieties that appear in Feynman integrals. While an increasingly large number of examples have now been identified to be Calabi-Yau, it remains unclear whether all such varieties have this property (and what this tells us about Feynman integrals in general).%
\footnote{The Calabi-Yau condition in the embedding we are considering restricts the degree of the defining polynomial; since we can deprojectivize and reprojectivize to increase the degree, it is effectively an upper bound. Thus, Calabi-Yaus are the first class one naturally encounters.}
In this paper, we have identified two further examples of Calabi-Yaus that can be realized as hypersurfaces in the weighted projective space $\mathbb{WP}^{1,\ldots,1,k}$ and have characterized hypersurfaces of this type in a number of ways. However, it again remains unclear how universal this property might be, and what it encodes about these specific Feynman graphs. To better connect the properties of these varieties to the physics encoded in Feynman diagrams, it may prove necessary to move to a differential equation approach~\cite{MullerStach:2012mp,Bloch:2014qca,Bloch:2016izu,Adams:2018yfj,mirrors_and_sunsets}.

There remains a great deal of technology to be developed before the integrals that we consider might be `computed'. It should be possible, for instance, to develop special functions analogous to the elliptic multiple polylogarithms~\cite{brown2011multiple,Adams:2017ejb,Broedel:2017kkb,Broedel:2017siw,Broedel:2018iwv,Broedel:2018qkq}, in terms of which these integrals could be evaluated. In particular, a coaction of the type that has proven useful in the polylogarithmic~\cite{Goncharov:2005sla,Brown:2011ik,Brown1102.1312,Duhr:2012fh} and elliptic cases~\cite{Broedel:2018iwv} should also exist for such functions~\cite{2015arXiv151206410B}. It should also be possible to develop iterated integral representations involving the relevant Calabi-Yau geometries, akin to what has been done for instance in refs.~\cite{brown2011multiple,Bogner:2019lfa}. Developing a better understanding of these spaces of functions is sure to lead to new surprises and simplifications, as has happened in the case of polylogarithmic and elliptic Feynman integrals over the last few years.

\vspace{\fill}\vspace{-4pt}
\section*{Acknowledgments}%
\vspace{-4pt}
We are grateful to Chuck Doran and Pierre Vanhove for useful discussions, and to Philip Candelas and Yinan Wang for involvement in earlier stages of this work.
We are also indebted to insightful comments from Gregory Korchemsky about the relevance of $\wheelint^{(4)}$ to the conformal fishnet theory, and to Miguel Paulos for discussions about Feynman parametrization and for encouraging us in particular to find a conformal parametric representation of $\wheelint^{(4)}$.
This work was supported in part by the Danish Independent Research Fund under grant number DFF-4002-00037 (MV, MW), the Danish National Research Foundation (DNRF91), the research grant 00015369 from Villum Fonden, a Starting Grant \mbox{(No.\ 757978)} from the European Research Council (JLB,AJM,MV,MvH,MW), the European Union's Horizon 2020 research and innovation program under grant agreement \mbox{No.\ 793151} (MvH), a Carlsberg Postdoctoral Fellowship (CF18-0641) (AJM), and the research grant 00025445 from Villum Fonden (MW).
Finally, JLB, AJM and MvH are grateful for the hospitality of the Aspen Center for Physics, which is supported by National Science Foundation grant PHY-1607611.

%================================================================================================================
\newpage
\appendix

\section{Desingularization by Complex Structure Deformation}
\label{sec:desing_cplx_structure}

The varieties we encounter when doing Feynman integrals are typically
singular; they may have singularities at fixed points in the Feynman
parameters or at points which vary with the external kinematics.  To
define a smooth variety, we deform the polynomial(s) that define the variety.  In
practice, this amounts to adding new monomials and changing the values
of the coefficients already present.  Such deformations turn out to be
\emph{complex structure} deformations.

One may worry that, even after performing such deformations, we do not
obtain a smooth variety.  At this point, we may invoke the Bertini
theorem (see for example ref.~\cite{Hubsch:1992nu} for a textbook
presentation).

\begin{theorem}[Bertini]
  Given a compact complex manifold \(X\) and a holomorphic line bundle
  \(L\) over \(X\) such that at every point \(x \in X\) the line bundle \(L\) has at
  least one non-zero section, then the points where a generic section
  \(f\) of \(L\) vanishes define a smooth hypersurface
  \(M = f^{-1}(0)\).
\end{theorem}

One way we can apply this theorem is to take the embedding space to be
\(\mathbb{P}^n\), and \(L\) to be a holomorphic line bundle
whose sections are homogeneous polynomials of degree \(d\).  Then the
Bertini theorem assures us that for a generic section \(f\) of \(L\)
i.e.\ for almost every choice of values for the coefficients of a
homogeneous degree \(d\) polynomial, the variety defined by
\(\{x \in \mathbb{P}^n \mid f(x) = 0\}\) is smooth.

In the following, we will apply reasoning analogous to the Bertini theorem
to embeddings in a weighted projective space of type
\(\mathbb{WP}^{1, \dots, 1, k}\).  Strictly speaking, the conditions
of the Bertini theorem are not satisfied since the embedding space
itself has a singularity.  If the singularity were to have dimension one or
larger, then it would generically intersect any hypersurface, and the
hypersurface would inherit the singularity.

However, in the case of \(\mathbb{WP}^{1, \dots, 1, k}\), the
singularity arises at just the point with homogeneous coordinates
\((0 , \dots , 0 , 1)\).  As a result, in the neighborhood of this point we
need to make the identifications
\begin{equation}
  (x_1 , \dots , x_k , 1) \simeq (\xi x_1 , \dots , \xi x_k , 1),
\end{equation}
where \(\xi\) is a \(k\)-th root of unity. Since the singularity arises at only a single point, a codimension-one
hypersurface will not generically contain it. (Moreover, we can
explicitly check to see if this happens.) In fact, even if our variety contains this singularity, we may define a
resolution and compute its Euler characteristic using for
example eqn.~(5.1.14) of ref.~\cite{Hubsch:1992nu}.

\section{Hodge Numbers and Euler Characteristic}\label{sec:hodge_numbers}

\subsection{Euler Characteristic}\label{sec:euler_characteristic}

One way to compute the Euler characteristic is to integrate the top
Chern class over the manifold.  We may obtain the Chern classes of an
embedded hypersurface from the Chern classes of the embedding manifold
and some data about the embedding.  There are several good
presentations of this material in the literature (see for example refs.~\cite{Hubsch:1992nu,Bouchard:2007ik}), so we will be brief.

Given a bundle \(E\), the total Chern class \(c(E)\) is the sum of all
Chern classes of all degrees.  Given an exact sequence of bundles
\(0 \to A \to B \to C \to 0\), we have \(c(B) = c(A) \wedge c(C)\). Using this fact, we conclude that the Chern class of a
weighted projective space with weights \((w_0, \dots, w_n)\) is
\begin{equation}
  c(\mathbb{WP}^{w_0, \dots, w_n}) = \prod_{i = 0}^n (1 + w_i J),
\end{equation}
where \(J = c_1(\mathcal{O}(1))\) is the first Chern class of the
bundle \(\mathcal{O}(1)\) whose sections are polynomials of
homogeneity one.  Depending on the weights \(w_i\), this bundle may
not exist as a holomorphic bundle on
\(\mathbb{WP}^{w_0, \dots, w_n}\), but can nevertheless be used as a
building block for other bundles.

We can define a codimension-\(m\) variety \(Y\) as the vanishing locus of \(m\) homogeneous polynomials of degrees \(d_i\), for \(i = 1, \dots, m\).  Then, the Chern class of \(Y\) is
\begin{equation}
  c(Y) = \frac {\prod_{i = 0}^n (1 + w_i J)}{\prod_{r = 1}^m (1 + d_r J)}.
\end{equation}
In this case, the Calabi-Yau condition reads
\begin{equation}
  \sum_{i = 0}^n w_i = \sum_{r = 1}^m d_r.
\end{equation}
Then the Euler characteristic is
\begin{equation}
  \chi(Y_{n-m}) = \int_Y c_{n - m} = \prod_{r=1}^m d_r \int_{\mathbb{WP}^{w_0, \dots, w_n}} c_{n - m} J^m,
\end{equation}
where we have extended the integral from \(Y\) to the full
\(\mathbb{WP}^{w_0, \dots, w_n}\) by wedging with a form that
encodes the contribution of the normal.

For our explicit examples of a codimension-one variety $X$ in $\mathbb{WP}^{1, \dots, 1, k}$, we have the Chern class
\begin{equation}
  c(X) = \frac {(1 + J)^k (1 + k J)}{1 + 2 k J},
\end{equation}
while the Euler characteristic is
\begin{equation}
  \chi(X_{k - 1}) = \int_X c_{k - 1}(X_{k - 1}) = \int_{\mathbb{WP}^{1, \dots, 1, k}} 2 k J \wedge c_{k - 1}(X_{k - 1}).
\end{equation}
The final piece of information we need is
\(\int_{\mathbb{WP}^{1, \dots, 1, k}} J^k = \frac 1 k\) because it
corresponds to the intersection of \(k\) hyperplanes at the singular
point \((0,\ldots,0,1)\), which has a cyclic singularity of order
\(k\).

Using this normalization, and the expression for \(c_{k - 1}\) obtained
by expanding the ratio of polynomials in \(J\),
 \begin{align}
 c_{k-1}(X_{k-1}) = \frac{1}{4 k} \left( 1 - (1 - 2 k)^k + 2 k^2 \right) J^{k-1} \, ,
 \end{align}
we eventually find
\begin{equation} \label{euler_characteristic}
  \chi(X_{k - 1}) = \frac {1 - (1 - 2 k)^k + 2 k^2}{2 k}.
\end{equation}
We have tabulated the Euler characteristic for the first few values of $k$ in table~\ref{tab:euler-characteristic-data}.

\subsection{Index Theorems}
\label{sec:index_theorems}

We can also compute further combinations of Hodge numbers as a cross-check using various
index theorems.  In particular, we have
\begin{gather}
  \chi(X_{k-1}) = \sum_r (-1)^r \dim H_{dR}^r(X) = \int_X c_{k-1}(X_{k-1}), \label{euler_characteristic_2}\\
  \chi_h(X_{k-1}) = \sum_q (-1)^q \dim H_{\bar{\partial}}^{0, q}(X_{k-1}) = \int_X td_{k-1}(X_{k-1}), \\
  \tau_H(X_{k-1}) = \sum_{p, q} (-1)^q \dim H_{\bar{\partial}}^{p, q}(X_{k-1}) = \int_X L_{k-1}(X_{k-1}),
\end{gather}
where \(\chi_h\) is the arithmetic genus and \(\tau_H\) is the
Hirzebruch signature.  Also, \(td\) is the Todd class and \(L\) is the
Hirzebruch polynomial. We present just the final answers for these computations:
\begin{align}
  k &= 3: \qquad \quad \chi_h = 2, \qquad \tau_H = -16, \\
  k &= 4: \qquad \quad \chi_h = 0, \qquad \tau_H = 0, \\
  k &= 5: \qquad \quad \chi_h = 2, \qquad \tau_H = 2002.
\end{align}
The reader can easily check that these values are consistent with the Hodge diamonds presented in section~\ref{sec:weighted-projective-space}.

\subsection{Lefschetz Hyperplane Theorem}\label{sec:lefschetz}

The cohomology of a hypersurface is strongly constrained by the
cohomology of the embedding space.  The Lefschetz-Bott theorem
characterizes the connections between these cohomology groups.  We
follow the presentations in ref.~\cite{Hubsch:1992nu} (see theorem 1.4 on
page 44).

In the Lefschetz-Bott theorem, we are given a complex compact manifold
\(X\) of dimension \(n + 1\) and a positive line bundle
\(\mathcal{L}\) over \(X\).  Then, given a holomorphic section
\(\lambda\), we denote by \(\lambda^{-1}(0)\) the points of \(X\)
where \(\lambda\) vanishes.  We then have\footnote{In
  fact, the result is more general and holds for homotopy groups.  The
  version for homology is listed as a corollary, presumably by an
  application of the Hurewicz theorem.}
\begin{align}
  H_q(\lambda^{-1}(0), \mathbb{Z}) &\simeq H_q(X, \mathbb{Z}), \qquad q \neq n, \label{Lefschetz_Bott_1} \\
  H_n(\lambda^{-1}(0), \mathbb{Z}) &\to H_n(X, \mathbb{Z}), \label{Lefschetz_Bott_2}
\end{align}
where the last map is surjective.  Dualizing to cohomology and using the Hodge
decomposition (and the fact that \((p, q)\)-forms pull back to
\((p, q)\)-forms), we obtain the result for cohomology. We can also use the Lefschetz-Bott theorem to constrain the cohomology
of complete intersections in projective spaces, by repeated
application of the theorem.

Stated concretely, equations~\eqref{Lefschetz_Bott_1} and~\eqref{Lefschetz_Bott_2} tell us that the upper and lower rows of the Hodge diamonds that describe
our Calabi-Yau hypersurfaces are inherited directly from $\mathbb{WP}^{1,\dots,1,k}$, while its middle row can involve numbers greater than or equal to those describing $\mathbb{WP}^{1,\dots,1,k}$. Interestingly, this means the Hodge numbers of these hypersurfaces could also arise from a codimension-one
embedding in unweighted projective space, which has Hodge numbers $h^{p,q}(\mathbb{P}^{k}) = \delta_{p,q}$. (We do not, however, know how to realize our Calabi-Yau hypersurfaces as embeddings in unweighted projective space.)

% \newpage
\vspace{-0pt}\section{Feynman Parametrization of the Three-Loop Wheel}\label{appendix:derivation_of_the_wheel}\vspace{-0pt}
% %%%%%%%%%%%%%%%%%%%%%%%%%%%%%%%%%%%%%%%%%%%%%%%%%%%%%%%%%%%%%%%%%%%%%%%%%%%

In this appendix, we describe the concrete steps by which the three-loop wheel
\eq{\wheelint^{(3)}\hspace{5pt}\bigger{\Leftrightarrow}\fig{-35.25pt}{1}{figures/wheel_integral_with_dual}\bigger{\Leftrightarrow}\fig{-35.25pt}{1}{figures/wheel_dual_with_integral}\,,\label{wheel_figure_appendix}}
defined in eqn.\ (\ref{wheel_spacetime_integral}) and discussed at length in \mbox{section \ref{sec:three-loop-wheel}}, can be expressed as a {\it rational} and {\it manifestly conformal} integral. This form was quoted in eqn.\ (\ref{eq:wheelint}).

Provided only a mild degree of cleverness, it is not hard to Feynman-parametrize and integrate each of the loop variables. This is especially true for (any choice of) the first two integrations, which are easily seen to be conformal box integrals. Let us briefly review the mechanics of how those integrals may be performed before applying these techniques to the integral in question.

\newpage\paragraph*{{\it Review}: Conformal Box Integrals in the Embedding Formalism}~\\[-12pt]

For the sake of reference and for those readers less familiar with the embedding formalism, let us recall that the box integral
\eq{\int\!\!\!d^4\!x_{{\color{hred}\ell}}\,\,\frac{1}{\x{{\color{hred}\ell}}{x_1}\x{{\color{hred}\ell}}{x_2}\x{{\color{hred}\ell}}{x_3}\x{{\color{hred}\ell}}{x_4}}\label{example_box}}
can be Feynman-parametrized by introducing
\eq{\embd{\mathcal{Y}}\equivR\alpha_1\embd{x_1}+\alpha_2\embd{x_2}+\alpha_3\embd{x_3}+\alpha_4\embd{x_4}}
so that the second Symanzik polynomial $\mathscr{F}$ may be written as
\eq{\mathscr{F}=\sum_{i\leq j}^4\alpha_i\alpha_j\x{x_i}{x_j}=\frac{1}{2}\x{\mathcal{Y}}{\mathcal{Y}}\equivL\xHalf{\mathcal{Y}}{\mathcal{Y}},\label{symanzik_def_of_prod}}
upon which the Feynman integral (\ref{example_box}) becomes
\eq{\int\!\!\!d^4\!x_{{\color{hred}\ell}}\,\,\frac{1}{\x{{\color{hred}\ell}}{x_1}\x{{\color{hred}\ell}}{x_2}\x{{\color{hred}\ell}}{x_3}\x{{\color{hred}\ell}}{x_4}}\propto\int\limits_0^\infty\!\!\!\proj{d^3\vec{\alpha}}\,\frac{1}{\xHalf{\mathcal{Y}}{\mathcal{Y}}^2}\,.\label{box_integral_in_ys}}
Above, we have used the notation $\proj{d^k\vec{\alpha}}$ to denote the volume form on $\mathbb{P}^k$ as expressed in terms of homogeneous coordinates $(\alpha_1,\alpha_2,\ldots,\alpha_{k+1})$. Specifically,
\eq{\proj{d^k\vec{\alpha}}\equivR d\alpha_1\cdots d{\alpha_{k+1}}\,\delta\big(\alpha_i-1\big)\label{proj_measure_defined}}
for any $\alpha_i$. The attentive reader will notice that Feynman's own de-projectivization prescription, $d\alpha_1\cdots d{\alpha_{k+1}}\,\delta\big(\sum_{i}\alpha_i-1\big)$, is related to that in eqn.\ (\ref{proj_measure_defined}) by a change of variables with unit Jacobian and which preserves the domain of integration, $\alpha_i\in[0,\infty]$.

Provided that there is at least one point $\embd{{\color{hblue}a_i}}$ such that $\x{a_i}{a_i}=0$, then $\xHalf{\mathcal{Y}}{\mathcal{Y}}$ will be linear in its Feynman parameter $\alpha_i$. When this happens, this Feynman parameter can be trivially integrated rationally. If the reader will forgive us for being somewhat pedantic, suppose that $\embd{\mathcal{Y}}$ may be written of the form
\eq{\embd{\mathcal{Y}}\equivL\embd{\mathcal{Q}}+\eta\embd{q}}
for any $\embd{q}$ such that $\x{q}{q}=0$ and for any $\eta\in\{\alpha_1,\ldots\}$; then
\eq{\xHalf{\mathcal{Y}}{\mathcal{Y}}=\xHalf{\mathcal{Q}}{\mathcal{Q}}+\eta\x{\mathcal{Q}}{q}\,,}
and
\eq{\int\limits_0^\infty\!\!\!\proj{d^3\vec{\alpha}}\,\frac{1}{\xHalf{\mathcal{Y}}{\mathcal{Y}}^2}=\int\limits_0^\infty\!\!\!\proj{d^2\vec{\alpha}}\int\limits_0^\infty\!\!\!d\eta\,\frac{1}{\big[\xHalf{\mathcal{Q}}{\mathcal{Q}}+\eta\x{\mathcal{Q}}{q}\big]^2}=\int\limits_0^\infty\!\!\!\proj{d^2\vec{\alpha}}\,\frac{1}{\xHalf{\mathcal{Q}}{\mathcal{Q}}\x{\mathcal{Q}}{q}}\,.}
The Feynman parametrization of the three-loop wheel integral follows directly from iteration of the above steps (with only mild cleverness at the end).

\paragraph*{The Feynman Parametrization of the Wheel Integral $\wheelint^{(3)}$}~\\[-12pt]

Let us begin with the (dual-momentum-)space-time definition of the wheel:
\eq{\wheelint^{(3)}\hspace{5pt}\equivR\int\!\!\!\frac{\fwboxR{0pt}{d^4\!x_{\!{\color{hred}A}}d^4\!x_{\!{\color{hred}B}}d^4\!x_{\hspace{-0.5pt}{\color{hred}C}}\,\hspace{6pt}}\x{a_1}{a_2}\hspace{-0.75pt}\x{b_1}{b_2}\hspace{-0.75pt}\x{c_1}{c_2}}{\x{{\color{hred}A}}{{\color{hred}C}}\x{{\color{hred}A}}{a_1}\hspace{-0.75pt}\x{{\color{hred}A}}{a_2}\hspace{-0.75pt}\x{{\color{hred}A}}{{\color{hred}B}}\hspace{-0.75pt}\x{{\color{hred}B}}{b_1}\hspace{-0.75pt}\x{{\color{hred}B}}{b_2}\hspace{-0.75pt}\x{{\color{hred}B}}{{\color{hred}C}}\hspace{-0.75pt}\x{{\color{hred}C}}{c_1}\hspace{-0.75pt}\x{{\color{hred}C}}{c_2}\hspace{-0.75pt}}\,.\label{wheel_int_appendix}}
We have used embedding-formalism-motivated notation to denote the squared-differences of points in dual-momentum space---i.e., $\x{a_1}{a_2}\equivR$\mbox{$({\color{hblue}\vec{a}_1}-{\color{hblue}\vec{a}_2})^2$}. Notice that all the points in dual-momentum space appearing in eqn.\ (\ref{wheel_int_appendix})---both those being integrated and those defining the external kinematics---satisfy $\x{x}{x}=0$.

Let us begin with the integration over the loop momentum $x_{{\color{hred}A}}$. It is not hard to see that this part of the integral is trivially identical to the box integral just discussed. Thus, we may introduce
\eq{\embd{\mathcal{Y}_{{\color{hred}A}}}\equivR\alpha_1\embd{a_1}+\alpha_2\embd{a_2}+\alpha_3\embd{{\color{hred}C}}+\eta_{\color{hred}A}\embd{{\color{hred}B}}\equivL\embd{\mathcal{Q}_{{\color{hred}A}}}+\eta_{\color{hred}A}\embd{{\color{hred}B}}\label{defn_of_ya_qa}}
and perform the integral over $x_{\color{hred}A}$ and $\eta_{\color{hred}A}$ to arrive at
\eq{\wheelint^{(3)}\hspace{2pt}=\int\limits_0^\infty\!\!\!\proj{d^2\!\vec{\alpha}}\int\!\!\!\frac{\fwboxR{0pt}{d^4\!x_{\!{\color{hred}B}}d^4\!x_{\hspace{-0.5pt}{\color{hred}C}}\,\hspace{6pt}}\x{a_1}{a_2}\hspace{-0.75pt}\x{b_1}{b_2}\hspace{-0.75pt}\x{c_1}{c_2}}{\xHalf{\mathcal{Q}_{{\color{hred}A}}}{\mathcal{Q}_{{\color{hred}A}}}\x{{\color{hred}B}}{\mathcal{Q}_{{\color{hred}A}}}\x{{\color{hred}B}}{b_1}\x{{\color{hred}B}}{b_2}\x{{\color{hred}B}}{{\color{hred}C}}\x{{\color{hred}C}}{c_1}\x{{\color{hred}C}}{c_2}}\,.\label{wheel_preantepenult}}

Now, as with $x_{\color{hred}A}$, the integral over $x_{{\color{hred}B}}$ in eqn. (\ref{wheel_preantepenult}) is just an ordinary conformal box integral. The only minor novelty is that one of the `propagators' of this integral, $\x{{\color{hred}B}}{\mathcal{Q}_{{\color{hred}A}}}$, involves a `non-simple' point in embedding space---one for which $\x{\mathcal{Q}_{{\color{hred}A}}}{\mathcal{Q}_{{\color{hred}A}}}\neq0$. This does not actually cause any trouble, however, because the Symanzik formalism defining the inner product $\x{\cdot}{\cdot}$ in eqn.\ (\ref{symanzik_def_of_prod}) did not require the points to be simple. (The simplicity of the external points only played a role in making it trivial to integrate out one Feynman parameter rationally.) Thus, we may introduce
\eq{\embd{\mathcal{Y}_{{\color{hred}B}}}\equivR\beta_1\embd{b_1}+\beta_2\embd{b_2}+\beta_3\embd{\mathcal{\mathcal{Q}}_{\color{hred}A}}+\eta_{\color{hred}B}\embd{{\color{hred}
C
% used to be B
}}\equivL\embd{\mathcal{Q}_{{\color{hred}B}}}+\eta_{\color{hred}B}\embd{{\color{hred}C}}}
and integrate over $x_{\color{hred}B}$ and $\eta_{\color{hred}B}$ to arrive at
\eq{\wheelint^{(3)}\hspace{2pt}=\int\limits_0^\infty\!\!\!\bigproj{d^2\!\vec{\alpha}}\!\!\!\bigproj{d^2\!\vec{\beta}}\int\!\!\!\frac{\fwboxR{0pt}{d^4\!x_{\hspace{-0.5pt}{\color{hred}C}}\,\hspace{6pt}}\x{a_1}{a_2}\hspace{-0.75pt}\x{b_1}{b_2}\hspace{-0.75pt}\x{c_1}{c_2}}{\xHalf{\mathcal{Q}_{{\color{hred}A}}}{\mathcal{Q}_{{\color{hred}A}}}\xHalf{\mathcal{Q}_{{\color{hred}B}}}{\mathcal{Q}_{{\color{hred}B}}}\x{{\color{hred}C}}{\mathcal{Q}_{{\color{hred}B}}}\x{{\color{hred}C}}{c_1}\x{{\color{hred}C}}{c_2}}\,.\label{wheel_antepenult}}

The careful reader should now be mildly worried as the integral over $x_{\color{hred}C}$ in eqn.\ (\ref{wheel_antepenult}) is not at all a recognizable (box) integral. Even worse: it is not even manifestly conformal in $x_{\color{hred}C}$! To appreciate the magnitude of this problem, notice that the factor $\x{\mathcal{Q}_{{\color{hred}A}}}{\mathcal{Q}_{{\color{hred}A}}}$ in the denominator of eqn.\ (\ref{wheel_antepenult}) involves a {\it sum} of terms with {\it different} conformal weights:
\eq{\xHalf{\mathcal{Q}_{{\color{hred}A}}}{\mathcal{Q}_{{\color{hred}A}}}=\alpha_1\alpha_2\x{a_1}{a_2}+\alpha_1\alpha_3\x{{\color{hred}C}}{a_1}+\alpha_2\alpha_3\x{{\color{hred}C}}{a_2}\,.}
%
% (Recall the definition of $\embd{\mathcal{Q}_{\color{hred}A}}$ in (\ref{defn_of_ya_qa}).)

Restoring conformality of this term turns out to be relatively easy. Consider rescaling the Feynman parameters $\alpha_i$ according to\footnote{We hope the reader can forgive the abuse of notation in using the same variables $\alpha_i$ to label the integration parameters before and after the rescaling.}
\eq{\alpha_1\mapsto\alpha_1\x{{\color{hred}C}}{a_2}\,,\quad\alpha_2\mapsto\alpha_2\x{{\color{hred}C}}{a_1}\,,\quad\alpha_3\mapsto\x{a_1}{a_2}\,.\label{wheel_alpha_change}}
Notice that we are actually eliminating the projective redundancy of $\proj{d^2\vec{\alpha}}$ by fixing $\alpha_3\mapsto\x{a_1}{a_2}$. (This is just done for notational compactness going forward.)

Under this rescaling,
\eq{\xHalf{\mathcal{Q}_{{\color{hred}A}}}{\mathcal{Q}_{{\color{hred}A}}}\underset{\text{(\ref{wheel_alpha_change})}}{\longmapsto}\x{a_1}{a_2}\x{{\color{hred}C}}{a_1}\x{{\color{hred}C}}{a_2}\big(\alpha_1+\alpha_2+\alpha_1\alpha_2\big)\,.\label{change_of_qa_in_wheel}}
The prefactor of eqn.\ \eqref{change_of_qa_in_wheel} cancels precisely against the Jacobian from eqn.\ (\ref{wheel_alpha_change}), resulting in
\eq{\wheelint^{(3)}\hspace{2pt}\underset{\text{(\ref{wheel_alpha_change})}}{\longmapsto}\int\limits_0^\infty\!\!\!d^2\!\vec{\alpha}\!\bigproj{d^2\!\vec{\beta}}\int\!\!\!\frac{\fwboxR{0pt}{d^4\!x_{\hspace{-0.5pt}{\color{hred}C}}\,\hspace{6pt}}\x{a_1}{a_2}\hspace{-0.75pt}\x{b_1}{b_2}\hspace{-0.75pt}\x{c_1}{c_2}}{(\alpha_1+\alpha_2+\alpha_1\alpha_2)\xHalf{\mathcal{Q}_{{\color{hred}B}}}{\mathcal{Q}_{{\color{hred}B}}}\x{{\color{hred}C}}{\mathcal{Q}_{{\color{hred}B}}}\x{{\color{hred}C}}{c_1}\x{{\color{hred}C}}{c_2}}\,.\label{wheel_antepenult_after_alpha_change}}

We have certainly improved the situation with respect to the $x_{\color{hred}C}$ integration, but not entirely. Notice, for example, that under the rescaling (\ref{wheel_alpha_change}), $\x{\mathcal{Q}_{{\color{hred}B}}}{\mathcal{Q}_{{\color{hred}B}}}$ becomes an irreducible (and inhomogeneous!) degree-two polynomial in $\embd{{\color{hred}C}}$. (This is trivial to see, considering eqn.\ (\ref{change_of_qa_in_wheel}), and $\x{\mathcal{Q}_{{\color{hred}B}}}{\mathcal{Q}_{{\color{hred}B}}}=\x{\mathcal{Q}_{{\color{hred}A}}}{\mathcal{Q}_{{\color{hred}A}}}+\ldots$\,.)

In fact, this problem can be remedied without too much hassle. Upon rescaling the $\beta_i$'s according to
\eq{\beta_1\mapsto\beta_1\frac{\x{{\color{hred}C}}{a_1}\x{a_1}{a_2}}{\x{a_1}{b_1}}\,,\quad\beta_2\mapsto\beta_2\frac{\x{{\color{hred}C}}{a_1}\x{a_1}{a_2}}{\x{a_1}{b_2}}\,,\quad\beta_3\mapsto1\,,\label{wheel_beta_change}}
and taking into account the corresponding Jacobian, the reader may verify that eqn.\ (\ref{wheel_antepenult_after_alpha_change}) takes the form
\eq{\wheelint^{(3)}\hspace{2pt}\underset{\text{(\ref{wheel_beta_change})}}{\longmapsto}\int\limits_0^\infty\!\!\!d^2\!\vec{\alpha}\,\,d^2\!\vec{\beta}\int\!\!\!\frac{\fwboxR{0pt}{d^4\!x_{\hspace{-0.5pt}{\color{hred}C}}\,\hspace{6pt}}\x{a_1}{a_2}^2\hspace{-0.75pt}\x{b_1}{b_2}\hspace{-0.75pt}\x{c_1}{c_2}/\x{a_1}{b_1}}{(\alpha_1+\alpha_2+\alpha_1\alpha_2)\x{{\color{hred}C}}{\mathcal{R}}\x{{\color{hred}C}}{\mathcal{S}}\x{{\color{hred}C}}{c_1}\x{{\color{hred}C}}{c_2}}\,,\label{wheel_penult}}
where we have defined the `propagators' $\x{{\color{hred}C}}{\mathcal{R}},\x{{\color{hred}C}}{\mathcal{S}}$ according to
\eq{\begin{split}
\embd{\mathcal{R}}&\equivR\phantom{+}\hspace{-4pt}\embd{a_2}(\alpha_1+\alpha_2)+\embd{b_1}\beta_1\frac{\x{a_1}{a_2}}{\x{a_1}{b_1}}+\embd{b_2}\beta_2\frac{\x{a_1}{a_2}}{\x{a_1}{b_2}}\,,\\
\embd{\mathcal{S}}&\equivR\phantom{+}\hspace{-4pt}\embd{\mathcal{R}}\x{a_1}{b_2}+\embd{a_1}\Big[\alpha_2\beta_1\frac{\x{a_1}{b_2}\x{a_2}{b_1}}{\x{a_1}{b_1}}+\alpha_2\beta_2\x{a_2}{b_2}+\beta_1\beta_2\frac{\x{a_1}{a_2}\x{b_1}{b_2}}{\x{a_1}{b_1}}\Big]\\
&\hspace{-4pt}\phantom{\equivR}\!+\!\!\embd{a_2}\Big[\alpha_1(\alpha_2+\beta_1+\beta_2)\x{a_1}{b_2}\Big]\,.
\end{split}}
Although these new propagators are not especially simple, we may now observe that eqn.\ (\ref{wheel_penult}) is a standard conformal box integral with respect to $x_{\color{hred}C}$(!). As such, our discussion above can be immediately applied. We merely introduce
\eq{\embd{\mathcal{Y}_{\color{hred}C}}\equivR\gamma_1\embd{c_1}+\gamma_2\embd{\mathcal{R}}+\gamma_3\embd{\mathcal{S}}+\eta_{\color{hred}C}\embd{c_2}\equivL\embd{\mathcal{Q}_{\color{hred}C}}+\eta_{\color{hred}C}\embd{c_2}\,,\label{yc_wheel_defined}}
and integrate over $x_{\color{hred}C}$ and $\eta_{\color{hred}C}$ to find
\eq{\wheelint^{(3)}\hspace{2pt}=\int\limits_0^\infty\!\!\!d^2\!\vec{\alpha}\,\,d^2\!\vec{\beta}\proj{d^2\!\vec{\gamma}}\,\,\frac{\x{a_1}{a_2}^2\hspace{-0.75pt}\x{b_1}{b_2}\hspace{-0.75pt}\x{c_1}{c_2}/\x{a_1}{b_1}}{(\alpha_1+\alpha_2+\alpha_1\alpha_2)\x{\mathcal{Q}_{\color{hred}C}}{c_2}\xHalf{\mathcal{Q}_{\color{hred}C}}{\mathcal{Q}_{\color{hred}C}}}\,.\label{wheel_ult1}}

We are essentially done. However, the representation (\ref{wheel_ult1}) is still not manifestly conformal in the external points. This can be quickly remedied. All we need to do is rescale the $\gamma_i$ Feynman parameters so that $\embd{\mathcal{Q}_{\color{hred}C}}$ in eqn.\ (\ref{yc_wheel_defined}) becomes uniform in weight. This can be achieved by rescaling them according to
\eq{\gamma_1\mapsto\gamma_1\frac{\x{a_1}{a_2}\x{a_2}{b_2}}{\x{a_2}{c_1}}\,,\quad\gamma_2\mapsto\gamma_2\x{a_1}{b_2}\,,\quad\gamma_3\mapsto1\,.\label{wheel_gamma_change}}
Upon including the Jacobian, gathering terms, and some minor simplifications, we obtain the formula quoted in eqn.\ (\ref{eq:wheelint})---namely, eqn.\ (\ref{wheel_ult1}) becomes
\eq{\wheelint^{(3)}\hspace{2pt}\underset{\text{(\ref{wheel_gamma_change})}}{\longmapsto}\int\limits_0^\infty\!\!\!d^2\!\vec{\alpha}\,\,d^2\!\vec{\beta}\,\,d^2\!\vec{\gamma}\,\,\frac{n_0}{f_1\,f_2\,f_3}\,,\label{wheel_ult}}
where
\vspace{-46pt}\eq{\begin{split}~\\[20pt]
n_0\equivR&\hspace{-4pt}\phantom{+}\xrv_1(\xru_1\xru_2\xru_3\xrv_1\xrv_2\xrv_3)\,,\\
f_1\equivR&\hspace{-4pt}\phantom{+}\alpha_1+\alpha_2+\alpha_1\alpha_2\,,\\
f_2\equivR&\hspace{-4pt}\phantom{+}\alpha_1(1+\alpha_2+\beta_1+\beta_2+\gamma_2)+\alpha_2(1+ \xru_1 \xrw_2(\xrw_3\beta_1+\beta_2)+\gamma_2)\\
&\hspace{-4pt}\!+\!\beta_1\xrv_1(1+ \xru_1\xru_3\xrv_2\xrw_2\beta_2+\gamma_2)+ \xru_2\xrv_1(\xru_1\xrv_3\gamma_1+\beta_2(1+\gamma_2))\,,\\
f_3\equivR&\fwboxL{0pt}{\hspace{-8pt}\phantom{+}(1\!+\!\alpha_2\!+\!\beta_1\!+\!\beta_2\!+\!\gamma_2)\Big[\alpha_1\hspace{-2pt}\Big(\hspace{-2pt}\gamma_1\!\!+\!\beta_2(1\!+\!\alpha_2\!+\! \xru_3\xrv_1\xrv_2\beta_1\!+\!\gamma_2)\!+\hspace{-1pt}\xrw_3\beta_1\hspace{-1pt}(1\!+\!\alpha_2\!+\!\hspace{-1pt}\gamma_2)\hspace{-2pt}\Big)}\\
&\hspace{-4pt}\!+\!(1\!+\!\hspace{-1pt}\gamma_2)\big(\hspace{-2pt}\xrw_3\alpha_2\beta_1\!+\!(\alpha_2\!+\!\xru_3 \xrv_1\xrv_2\beta_1)\beta_2\hspace{-2pt}\big)\hspace{-2pt}\Big]\!+\!\gamma_1\Big[\alpha_2(1\!+\!\xru_1(\xrw_3\beta_1\!+\!\beta_2)\!\!+\!\!\gamma_2)\\
&\hspace{-4pt}\!+\!\xru_3 \xrv_1(\xru_2 \xrw_1\beta_2(1\!+\!\gamma_2)\!+\!\beta_1(1\!+\!\xru_1 \xrv_2\beta_2\!+\!\gamma_2))\Big]\,,\\[-26pt]\label{wheel_int_den_defns}
\end{split}
}
expressed in terms of the basis of dual-conformal invariant cross-ratios
\eq{\hspace{-70pt}\begin{array}{@{}r@{}l@{$\hspace{15pt}$}r@{}l@{$\hspace{15pt}$}r@{}l@{}}\xru_1\!\equivR&\!\u{c_1a_1}{a_2b_2}\,,&\xru_2\!\equivR&\!\u{a_1b_1}{b_2c_2}\,,&\xru_3\!\equivR&\!\u{b_1c_1}{c_2a_2}\,,\\
\xrv_1\!\equivR&\!\u{a_1a_2}{b_1c_2}\,,&\xrv_2\!\equivR&\!\u{b_1b_2}{c_1a_2}\,,&\xrv_3\!\equivR&\!\u{c_1c_2}{a_1b_2}\,,\\
\xrw_1\!\equivR&\!\u{b_2c_1}{c_2b_1}\,,&\xrw_2\!\equivR&\!\u{c_2a_1}{a_2c_1}\,,&\xrw_3\!\equivR&\!\u{a_2b_1}{b_2a_1}\,.\end{array}\hspace{-50pt}\label{wheel_cross_ratios_defn}}\vspace{-10pt}
Recall that these are defined according to
\vspace{5pt}\eq{\u{{\color{hgreen}x}{\color{hred}y}}{{\color{hblue}z}{\color{hgreen}w}}\equivR\frac{\x{{\color{hgreen}x}}{{\color{hred}y}}\x{{\color{hblue}z}}{{\color{hgreen}w}}}{\x{{\color{hgreen}x}}{{\color{hblue}z}}\x{{\color{hred}y}}{{\color{hgreen}w}}}\,.}

Although there appeared to be some magic in the Feynman-parametric rescaling in eqn.\ (\ref{wheel_beta_change})---which restored not only conformality in the $x_{{\color{hred}C}}$ integration, but also its manifest linearity in each factor of the denominator of eqn.\ (\ref{wheel_penult})---this magic in some sense `had to work'. Indeed, Miguel Paulos has shown \cite{MiguelPaulosprivate} that all dual-conformal Feynman integrals whose dual-graphs involve internal loop momenta connected via trees are always possible to compute conformally by integrating one loop at a time (as described in \mbox{ref.\ \cite{Bourjaily:2019jrk}}) and rescaling Feynman parameters accordingly. His proof extends also to integrals whose dual graphs are free of four-cycles---and hence, his argument also applies to $\wheelint^{(3)}$. Nevertheless, the existence of four-cycles in the dual graph (as will be the case for $\wheelint^{(4)}$ discussed below) prevent this line of reasoning from being applied. As such, it is natural to wonder if there is any obstruction to the magic found in the rescaling (\ref{wheel_beta_change}) when considered in the context of a four (or higher-)loop wheel.

% \newpage
\vspace{-0pt}\section{Feynman Parametrization of the Four-Loop Wheel}\label{appendix:derivation_of_the_wheel_four}\vspace{-0pt}

Similarly to three loops, the four-loop wheel (also known as the `window' integral) %$\wheelint^{(4)}$
can be defined in dual-momentum space as
\eq{\hspace{-28pt}\fwbox{0pt}{\hspace{0pt}\wheelint^{(4)}\hspace{-0pt}\equivR\!\!\int\!\!\!\frac{\fwboxR{0pt}{d^4\!x_{\!{\color{hred}A}}d^4\!x_{\!{\color{hred}B}}d^4\!x_{\hspace{-0.5pt}{\color{hred}C}}d^4\!x_{\hspace{-0.5pt}{\color{hred}D}}\,\hspace{10pt}}\x{a_1}{a_2}\hspace{-0.75pt}\x{b_1}{b_2}\hspace{-0.75pt}\x{c_1}{c_2}\x{d_1}{d_2}}{\x{{\color{hred}D}}{{\color{hred}A}}\x{{\color{hred}A}}{a_1}\hspace{-0.75pt}\x{{\color{hred}A}}{a_2}\hspace{-0.75pt}\x{{\color{hred}A}}{{\color{hred}B}}\hspace{-0.75pt}\x{{\color{hred}B}}{b_1}\hspace{-0.75pt}\x{{\color{hred}B}}{b_2}\hspace{-0.75pt}\x{{\color{hred}B}}{{\color{hred}C}}\hspace{-0.75pt}\x{{\color{hred}C}}{c_1}\hspace{-0.75pt}\x{{\color{hred}C}}{c_2}\hspace{-0.75pt}\x{{\color{hred}C}}{{\color{hred}D}}\hspace{-0.75pt}\x{{\color{hred}D}}{d_1}\hspace{-0.75pt}\x{{\color{hred}D}}{d_2}\hspace{-0.75pt}}\,.}\label{window_int_appendix}}
As before, sequentially introducing Feynman parameters will proceed semi-trivially until the last step as each integral is a standard, conformal box integral. Thus, we may save ourselves some of the pedantry of the previous discussion and cut to the chase---to the non-trivial steps at the end.

To integrate over the first three loop momenta, $x_{\color{hred}A},x_{\color{hred}B},x_{\color{hred}C}$ in eqn.\ (\ref{window_int_appendix}), we introduce Feynman parameters according to
\eq{\begin{array}{@{}l@{$\equivR$}l@{$\,+\,$}c@{$\,+\,$}c@{$\,+\,$}r@{$\equivL$}l@{$\,+\,$}l@{$\,,$}}
\embd{\mathcal{Y}_{\color{hred}A}}&\alpha_1\embd{a_1}&\alpha_2\embd{a_2}&\alpha_3\embd{{\color{hred}D}}&\eta_{\color{hred}A}\embd{{\color{hred}B}}&\embd{\mathcal{Q}_{\color{hred}A}}&\eta_{\color{hred}A}\embd{{\color{hred}B}}\\
\embd{\mathcal{Y}_{\color{hred}B}}&\beta_1\embd{b_1}&\beta_2\embd{b_2}&\beta_3\embd{\mathcal{Q}_{\color{hred}A}}&\eta_{\color{hred}B}\embd{{\color{hred}C}}&\embd{\mathcal{Q}_{\color{hred}B}}&\eta_{\color{hred}B}\embd{{\color{hred}C}}\\
\embd{\mathcal{Y}_{\color{hred}C}}&\gamma_1\embd{c_1}&\gamma_2\embd{c_2}&\gamma_3\embd{\mathcal{Q}_{\color{hred}B}}&\eta_{\color{hred}C}\embd{{\color{hred}D}}&\embd{\mathcal{Q}_{\color{hred}C}}&\eta_{\color{hred}C}\embd{{\color{hred}D}}
\end{array}\label{window_ys_defn}}
and integrate over the Feynman parameters $\eta_{\color{hred}A},\eta_{\color{hred}B},\eta_{\color{hred}C}$ to arrive at
\eq{\hspace{-28pt}\fwbox{0pt}{\hspace{0pt}\wheelint^{(4)}\hspace{-0pt}=\!\!\int\limits_0^\infty\!\!\!\bigproj{d^2\!\vec{\alpha}}\!\!\!\bigproj{d^2\!\vec{\beta}}\!\!\!\bigproj{d^2\!\vec{\gamma}}\int\!\!\!\frac{\fwboxR{0pt}{d^4\!x_{\hspace{-0.5pt}{\color{hred}D}}\,\hspace{18pt}}\x{a_1}{a_2}\hspace{-0.75pt}\x{b_1}{b_2}\hspace{-0.75pt}\x{c_1}{c_2}\x{d_1}{d_2}}{\xHalf{\mathcal{Q}_{\color{hred}A}}{\mathcal{Q}_{\color{hred}A}}\xHalf{\mathcal{Q}_{\color{hred}B}}{\mathcal{Q}_{\color{hred}B}}\xHalf{\mathcal{Q}_{\color{hred}C}}{\mathcal{Q}_{\color{hred}C}}\x{{\color{hred}D}}{\mathcal{Q}_{\color{hred}C}}\x{{\color{hred}D}}{d_1}\hspace{-0pt}\x{{\color{hred}D}}{d_2}}\,.}\label{window_int_d_pre_rescaling}}
As was the case with three loops, we now find an obstruction in the last loop integration of eqn.\ (\ref{window_int_d_pre_rescaling}), as it is far from manifestly conformal.

(To reiterate a point made above, we should be clear that {\it mere} conformality is not sufficient for us to Feynman parametrize and do the loop integrations. For example, consider an integral of the form
\eq{\int\!\!\!d^4\!x_{{\color{hred}\ell}}\,\,\frac{1}{\x{{\color{hred}\ell}}{a}\x{{\color{hred}\ell}}{b}\big[\x{{\color{hred}\ell}}{c}\x{{\color{hred}\ell}}{d}+\x{{\color{hred}\ell}}{e}\x{{\color{hred}\ell}}{f}\big]}\,.\label{example_bad_box}}
We know of no method by which such integrals can be systematically integrated.\footnote{Integrands such as (\ref{example_bad_box}) arise in the context of all-loop recursion relations \cite{ArkaniHamed:2010kv}, and it would be incredibly interesting to develop methods for these integrations.} In this work, we take a much more conservative approach, and demand that integrands be brought to the form such that their (loop-dependent) denominators are built directly as products of propagators.)

Somewhat surprisingly, it turns out to be fairly straightforward to bring eqn.\ (\ref{window_int_d_pre_rescaling}) into a recognizable form by a sequence of rescalings as done for three loops. In particular, if we rescale (and eliminate the projective redundancy of) the Feynman parameters according to\footnote{A more symmetrical choice of rescalings---one which treats the $\gamma_i$'s more similarly to the $\beta_i$'s---would have worked. We have chosen the somewhat unbalanced set of rescalings in order to maximize the number of smoothly accessible toy-model-like limits.}
%@{
\eq{\begin{array}{@{}r@{$\,$}c@{$\,$}l@{$\hspace{20pt}$}r@{$\,$}c@{$\,$}l@{$\hspace{20pt}$}r@{$\,$}c@{$\,$}l@{}c@{}}
\alpha_1&\mapsto&\alpha_1\x{{\color{hred}D}}{a_2},&\alpha_2&\mapsto&\alpha_2\x{{\color{hred}D}}{a_1},&\alpha_3&\mapsto&1\!\times\!\x{a_1}{a_2},\,\\
\beta_1&\mapsto&\beta_1\displaystyle\frac{\x{{\color{hred}D}}{a_1}\x{a_1}{a_2}}{\x{a_1}{b_1}},&\beta_2&\mapsto&\beta_2\displaystyle\frac{\x{{\color{hred}D}}{a_1}\x{a_1}{a_2}}{\x{a_1}{b_2}},&\beta_3&\mapsto&1,\,\\
\gamma_1&\mapsto&\gamma_1\displaystyle\frac{\x{{\color{hred}D}}{a_1}\x{a_1}{a_2}\x{b_1}{b_2}}{\x{a_1}{b_2}\x{b_1}{c_1}},&\gamma_2&\mapsto&\gamma_2\displaystyle\frac{\x{{\color{hred}D}}{a_1}\x{a_1}{a_2}\x{b_1}{b_2}}{\x{a_1}{b_1}\x{b_2}{c_2}},&\gamma_3&\mapsto&1,\,\\
\end{array}\label{window_abc_rescaling_rules}}
then the integral (\ref{window_int_d_pre_rescaling}) becomes
\eq{\hspace{-28pt}\fwbox{0pt}{\hspace{0pt}\wheelint^{(4)}\hspace{-0pt}\underset{\text{(\ref{window_abc_rescaling_rules})}}{\longmapsto}\int\limits_0^\infty\!\!\!d^2\!\vec{\alpha}\,\,d^2\!\vec{\beta}\,\,d^2\!\vec{\gamma}\int\!\!\!\frac{\fwboxR{0pt}{d^4\!x_{\hspace{-0.5pt}{\color{hred}D}}\,\hspace{28pt}}\kappa\,\x{{\color{hred}D}}{a_1}}{(\alpha_1+\alpha_2+\alpha_1\alpha_2)\x{{\color{hred}D}}{\mathcal{R}}\x{{\color{hred}D}}{\mathcal{S}}\x{{\color{hred}D}}{\mathcal{T}}\x{{\color{hred}D}}{d_1}\hspace{-0pt}\x{{\color{hred}D}}{d_2}}\,,}\label{window_int_d_post_rescaling}}
where the prefactor in the numerator is
\eq{\kappa\equivR\frac{\x{a_1}{a_2}^3\x{b_1}{b_2}^3\x{c_1}{c_2}\x{d_1}{d_2}}{\x{a_1}{b_1}^2\x{a_1}{b_2}^2\x{b_1}{c_1}\x{b_2}{c_2}},}
which arises from the various Jacobians. Moreover, the new `propagators' are
\eq{\begin{split}
\embd{\mathcal{R}}&\equivR\phantom{+}\hspace{-4pt}\embd{a_1}\Big[\alpha_2\beta_1\frac{\x{a_2}{b_1}}{\x{a_1}{b_1}}+\alpha_2\beta_2\frac{\x{a_2}{b_2}}{\x{a_1}{b_2}}+\beta_1\beta_2\frac{\x{a_1}{a_2}\x{b_1}{b_2}}{\x{a_1}{b_1}\x{a_1}{b_2}}\Big]\\
&\hspace{-4pt}\phantom{\equivR}\!+\!\!\embd{a_2}\big[\alpha_1(1+\alpha_2+\beta_1+\beta_2)+\alpha_2\big]+\embd{\mathcal{U}}\,,\\
\embd{\mathcal{S}}&\equivR\phantom{+}\hspace{-4pt}\embd{a_2}(\alpha_1+\alpha_2)+\embd{\mathcal{U}}+\embd{\mathcal{V}}\,,\\
\embd{\mathcal{T}}&\equivR\phantom{+}\hspace{-4pt}\embd{\mathcal{S}}+\embd{a_1}f_{\mathcal{T}}+\embd{a_2}\alpha_1\Big[\alpha_2+\beta_1+\beta_2+\gamma_1\frac{\x{a_1}{c_1}\x{b_1}{b_2}}{\x{a_1}{b_2}\x{b_1}{c_1}}+\gamma_2\frac{\x{a_1}{c_2}\x{b_1}{b_2}}{\x{a_1}{b_1}\x{b_2}{c_2}}\Big]\,,\\
\end{split}\label{window_rst_defined}}
in terms of
\eq{\begin{split}
\embd{\mathcal{U}}&\equivR\phantom{+}\hspace{-4pt}\embd{b_1}\beta_1\frac{\x{a_1}{a_2}}{\x{a_1}{b_1}}+\embd{b_2}\beta_2\frac{\x{a_1}{a_2}}{\x{a_1}{b_2}}\,,\\
\embd{\mathcal{V}}&\equivR\phantom{+}\hspace{-4pt}\embd{c_1}\gamma_1\frac{\x{a_1}{a_2}\x{b_1}{b_2}}{\x{a_1}{b_2}\x{b_1}{c_1}}+\embd{c_2}\gamma_2\frac{\x{a_1}{a_2}\x{b_1}{b_2}}{\x{a_1}{b_1}\x{b_2}{c_2}}\,,
\end{split}}
and where we have defined the scalar function
\eq{\begin{split}
f_{\mathcal{T}}\equivR&\hspace{-2pt}\frac{\x{a_1}{a_2}\x{b_1}{b_2}}{\x{a_1}{b_1}\x{a_1}{b_2}}\left[\beta_1\beta_2+\beta_1\gamma_1+\beta_2\gamma_2+\alpha_2\left(\beta_1\frac{\x{a_1}{b_2}\x{a_2}{b_1}}{\x{a_1}{a_2}\x{b_1}{b_2}}\right.\right.\\
&\hspace{-4pt}+\beta_2\frac{\x{a_1}{b_1}\x{a_2}{b_2}}{\x{a_1}{a_2}\x{b_1}{b_2}}+\gamma_1\frac{\x{a_1}{b_1}\x{a_2}{c_1}}{\x{a_1}{a_2}\x{b_1}{c_1}}+\left.\gamma_2\frac{\x{a_1}{b_2}\x{a_2}{c_2}}{\x{a_1}{a_2}\x{b_2}{c_2}}\right)\\
&\hspace{-4pt}+\beta_1\gamma_2\frac{\x{a_1}{b_2}\x{b_1}{c_2}}{\x{a_1}{b_1}\x{b_2}{c_2}}+\beta_2\gamma_1\frac{\x{a_1}{b_1}\x{b_2}{c_1}}{\x{a_1}{b_2}\x{b_1}{c_1}}+\left.\gamma_1\gamma_2\frac{\x{b_1}{b_2}\x{c_1}{c_2}}{\x{b_1}{c_1}\x{b_2}{c_2}}\right]\,.
\end{split}}
The integral (\ref{window_int_d_post_rescaling}) is a conformal integral (with respect to $x_{\color{hred}D}$) which can be done almost as trivially as the box integral. In particular, its Feynman parametrization follows more-or-less trivially from differentiation (with respect to ${\color{hred}\ell}$) of the (Feynman parametrized) box integral. (The interested reader should consult, e.g., ref.\ \cite{Bourjaily:2019jrk}.)

Feynman parametrization of the integral (\ref{window_int_d_post_rescaling}) may be done by introducing
\eq{\embd{\mathcal{Y}_{\color{hred}D}}\equivR\delta_1\embd{d_1}+\delta_2\embd{\mathcal{R}}+\delta_3\embd{\mathcal{S}}+\delta_4\embd{\mathcal{T}}+\eta_{\color{hred}D}\embd{d_2}\equivL\embd{\mathcal{Q}_{\color{hred}D}}+\eta_{\color{hred}D}\embd{d_2}\,,\label{defn_of_final_y_for_window}}
and integrating over $x_{\color{hred}D}$ in the ordinary way. This results in a representation of $\wheelint^{(4)}$ of the form
\begin{align}
\wheelint^{(4)}\hspace{-0pt}&=\int\limits_0^\infty\!\!\!d^2\!\vec{\alpha}\,\,d^2\!\vec{\beta}\,\,d^2\!\vec{\gamma}\bigproj{d^3\!\vec{\delta}}\int\limits_0^\infty\!\!\!d\eta_{\color{hred}D}\frac{\kappa\,\x{\mathcal{Y}_{\color{hred}D}}{a_1}}{(\alpha_1+\alpha_2+\alpha_1\alpha_2)\x{\mathcal{Y}_{\color{hred}D}}{\mathcal{Y}_{\color{hred}D}}^3}\nonumber\\
&=\int\limits_0^\infty\!\!\!d^2\!\vec{\alpha}\,\,d^2\!\vec{\beta}\,\,d^2\!\vec{\gamma}\bigproj{d^3\!\vec{\delta}}\int\limits_0^\infty\!\!\!d\eta_{\color{hred}D}\frac{\kappa\,\big(\x{\mathcal{Q}_{\color{hred}D}}{a_1}+\eta_{\color{hred}D}\x{d_2}{a_1}\big)}{(\alpha_1+\alpha_2+\alpha_1\alpha_2)\big[\x{\mathcal{Q}_{\color{hred}D}}{\mathcal{Q}_{\color{hred}D}}+\eta_{\color{hred}D}\x{\mathcal{Q}_{\color{hred}D}}{d_2}\big]^3}\label{window_penult_integration_steps}\\
&=\int\limits_0^\infty\!\!\!d^2\!\vec{\alpha}\,\,d^2\!\vec{\beta}\,\,d^2\!\vec{\gamma}\bigproj{d^3\!\vec{\delta}}\frac{\kappa}{2(\alpha_1\!+\!\alpha_2\!+\!\alpha_1\alpha_2)}\left[\frac{\x{d_2}{a_2}}{\x{\mathcal{Q}_{\color{hred}D}}{d_2}^2\x{\mathcal{Q}_{\color{hred}D}}{\mathcal{Q}_{\color{hred}D}}}\!+\!\frac{\x{\mathcal{Q}_{\color{hred}D}}{a_1}}{\x{\mathcal{Q}_{\color{hred}D}}{d_2}\x{\mathcal{Q}_{\color{hred}D}}{\mathcal{Q}_{\color{hred}D}}^2}
\right].\nonumber
\end{align}

As before, the only thing we must do to render the expression (\ref{window_penult_integration_steps}) manifestly conformal with respect to the external momenta is to rescale the $\delta_i$'s such that $\embd{\mathcal{Q}_{\color{hred}D}}$ becomes uniform in weight. This is in fact easy, as the reader can easily observe that all of the factors defined in eqn.\ (\ref{window_rst_defined}) scale like $\embd{a_2}$; as such, the only term in eqn.\ (\ref{defn_of_final_y_for_window}) which has the wrong scaling weights is the first one. Rescaling as required and eliminating the projective redundancy (now just for consistency with the previous analysis) according to
\eq{\delta_1\mapsto\delta_1\x{a_1}{a_2}\x{a_1}{d_1}\,,\qquad\delta_4\mapsto1\,,\label{final_rescaling_of_window}}
the four-loop wheel takes the form
\eq{\wheelint^{(4)}\hspace{-0pt}\underset{\text{(\ref{final_rescaling_of_window})}}{\longmapsto}\int\limits_0^\infty\!\!\!d^2\!\vec{\alpha}\,\,d^2\!\vec{\beta}\,\,d^2\!\vec{\gamma}\,\,d^3\!\vec{\delta}\frac{n_0}{f_1\,f_2\,f_3}\left(\frac{n_1}{f_2}+\frac{n_2}{f_3}\right)\,,\label{ult_form_of_window}}
where $n_i$'s and the $f_i$'s are all {\it directly} expressible in terms of dual-conformally invariant cross-ratios.

We might ask if we could have done better, and found a representation as an eight-fold integral. The difficulty here is in dealing with the final pentagon integral, which we here represent as a three-fold. These integrals can be expanded into boxes, and this would indeed give rise to a two-fold representation. However, writing out this box expansion shows that it contains dilogs which have square-root arguments---and these square roots would involve the other Feynman parameters. As such, while one can indeed write down some two-fold representation, it would not help us to understand its transcendental properties. At present, we know of no way to write the four-loop wheel as a rational eight-fold integral.

\subsection{\texorpdfstring{Interesting Kinematic Limits of the Wheel Integral $\wheelint^{(4)}$}{Interesting Kinematic Limits of the Four-Loop Wheel Integral}}
\label{subsec:window_limits}

The four-loop wheel integral has several interesting kinematic limits. We discuss them below, and provide expressions for the integral in each of these limits in \textsc{Mathematica} format in the .ancillary file \texttt{integrands\_and\_varieties.m}.

\paragraph*{The `Fishnet' Limit of the Wheel Integral $\wheelint^{(4)}$}~\\[-12pt]

The first limit we consider is the one in which all middle legs are light-like:
\eq{\x{a_2}{b_1}=\x{b_2}{c_1}=\x{c_2}{d_1}=\x{d_2}{a_1}=0\,.\label{window_fishnet_limit}}

\vspace{-10pt}\eq{\fig{-34.9pt}{1}{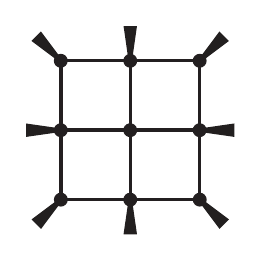}\bigger{\underset{\text{(\ref{window_fishnet_limit})}}{\Longrightarrow}}\fig{-34.9pt}{1}{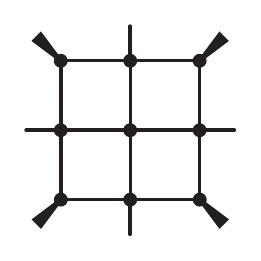}\bigger{\Leftrightarrow}\fig{-34.9pt}{1}{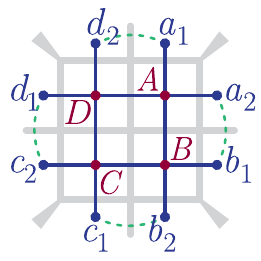}\label{window_massless_middle_limit}\vspace{-0pt}}
Notice that a particular case of this limit---where the `massive' momenta flowing into the corners of the wheel are pairs of massless particles---is itself a particular planar amplitude in the integrable conformal fishnet theory \mbox{\cite{Gurdogan:2015csr,Sieg:2016vap,Grabner:2017pgm}},
\eq{\mathcal{A}({\color{hblue}\varphi_{12}},{\color{hblue}\varphi_{12}},{\color{hblue}\varphi_{12}},{\color{hred}\varphi_{13}},{\color{hred}\varphi_{13}},{\color{hred}\varphi_{13}},
{\color{hblue}\varphi_{34}},{\color{hblue}\varphi_{34}},{\color{hblue}\varphi_{34}},{\color{hred}\varphi_{24}},{\color{hred}\varphi_{24}},{\color{hred}\varphi_{24}})=\fig{-34.9pt}{1}{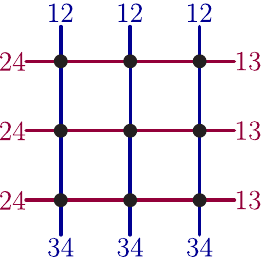}\,\,\,,}
which is also a particular \emph{component amplitude} of the 12-point N$^4$MHV scattering amplitude in planar $\mathcal{N}=4$ supersymmetric Yang-Mills theory, $\mathcal{A}_{12}^{(4)}$. This component of the supersymmetric amplitude corresponds to
\eq{\int\!\!\!\big(d\tilde{\eta}^1_1d\widetilde{\eta}^1_2\cdots d\widetilde{\eta}^1_6\big)\big(d\widetilde{\eta}^2_{10}d\widetilde{\eta}^2_{11}\cdots d\widetilde{\eta}^2_3\big)\big(d\widetilde{\eta}^3_{4}d\widetilde{\eta}^3_{5}\cdots d\widetilde{\eta}^3_9\big)\big(d\widetilde{\eta}^4_{7}d\widetilde{\eta}^4_{8}\cdots d\widetilde{\eta}^4_{12}\big)\mathcal{A}_{12}^{(4)}\,.}
We also note that in this limit (and hence all those below it), $n_1$ of eqn.\ (\ref{ult_form_of_window}) vanishes.

\paragraph*{A Nine-Dimensional Toy Model of the Wheel Integral $\wheelint^{(4)}$}~\\[-12pt]

This limit is analogous to the toy models discussed in section~\ref{sec:three-loop-wheel} and ref.~\cite{Bourjaily:2017bsb}.
In this case, there are several ways to `route' 8 light-like points among the external points. The only one which will be dihedrally invariant is the one defined by the conditions (\ref{window_fishnet_limit}) and
\eq{\x{a_1}{c_2}=\x{a_2}{c_1}=\x{b_1}{d_2}=\x{b_2}{d_1}\,.\label{nine_parameter_toy_model_limit_of_window}}

\vspace{-10pt}\eq{\fig{-34.9pt}{1}{figures/window_integral_degn1}\bigger{\Leftrightarrow}\fig{-34.9pt}{1}{figures/dual_window_degn1}\bigger{\underset{\text{(\ref{nine_parameter_toy_model_limit_of_window})}}{\Longrightarrow}}\fig{-34.9pt}{1}{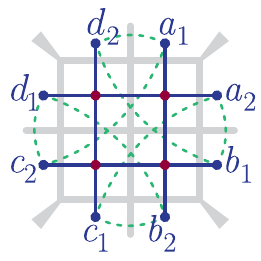}\label{window_generic_toy_model_limit}\vspace{-0pt}}
In this limit, the integral will depend on the space of kinematics associated with 8 pairwise light-like separated points---a nine-dimensional parameter space. We do not expect this limit to lead to any drop in rigidity.

\paragraph*{The Basso-Dixon Fishnet Integral $I_{2,2}$ as a Limit of $\wheelint^{(4)}$}~\\[-12pt]

Another special case of interest is the Basso-Dixon fishnet integral $I_{2,2}$, which contributes to the four-point correlation function in planar $\varphi^4$ theory. This corresponds to taking the limit where the eight dual points defining the wheel integral $\wheelint^{(4)}$ are pairwise identified according to
\eq{d_2=a_1,\quad a_2=b_1,\quad b_2=c_1,\quad c_2=d_1\,.\label{basso_dixon_limit_of_window}}
Graphically, this corresponds to
\vspace{-10pt}\eq{\fig{-34.9pt}{1}{figures/window_integral_plain}\bigger{\Leftrightarrow}\fig{-34.9pt}{1}{figures/window_dual_with_integral}\bigger{\underset{\text{(\ref{basso_dixon_limit_of_window})}}{\Longrightarrow}}\fig{-34.9pt}{1}{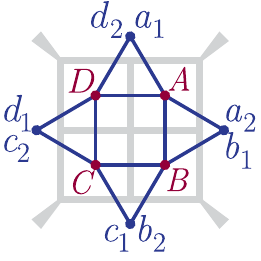}\bigger{\Leftrightarrow}\fig{-34.9pt}{1}{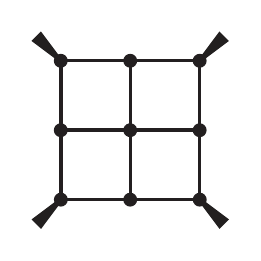}\!.\label{window_basso_dixon_limit}\vspace{-0pt}}
This limit is known explicitly \cite{Basso:2017jwq}, and in particular is polylogarithmic.

\paragraph*{A Two-Dimensional Toy Model of the Wheel Integral $\wheelint^{(4)}$}~\\[-12pt]

One final limit of interest is one that appeared in \mbox{ref. \cite{Gromov:2018hut}}---also in the context of the conformal fishnet theory. This limit corresponds to a different pairwise identification of the eight dual points which define the integral, namely,
\eq{a_1=c_2,\quad a_2=c_1,\quad b_1=d_2,\quad b_2=d_1\,.\label{window_2magnon_limit_defn}}
This limit can perhaps be best understood as a `non-planar' gluing of the original dual integral---obtained via the sequence
\eq{\fwbox{0pt}{\hspace{-28pt}\fig{-34.9pt}{1}{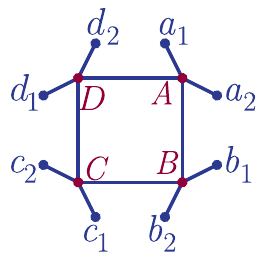}\bigger{\simeq}\hspace{-4pt}\fig{-34.9pt}{1}{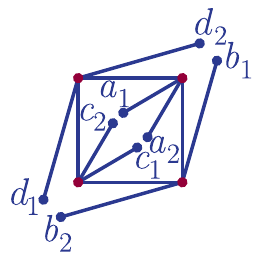}\hspace{-6pt}\bigger{\underset{\text{(\ref{window_2magnon_limit_defn})}}{\Longrightarrow}}\fig{-34.9pt}{1}{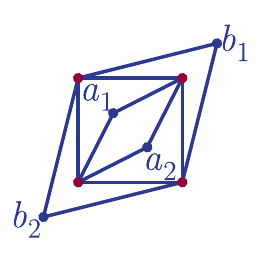}\hspace{-4pt}\bigger{\simeq}\fig{-21.9pt}{1}{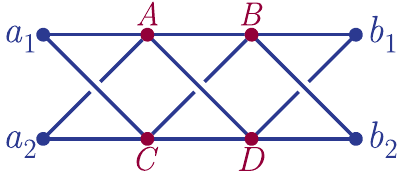}.}\label{window_to_2magnon_limit_figures}}
In this limit, the integral can be seen to contribute to the `2-magnon' 4-point function as drawn on the right-hand part of figure 1 of \mbox{ref. \cite{Gromov:2018hut}}. At leading order, this four-point function is given by a single Feynman integral: that drawn in eqn.\ (\ref{window_to_2magnon_limit_figures}). This function is known to be non-polylogarithmic. Fourier-transformed, it corresponds to the five-loop amoeba integral of ref.~\cite{Bourjaily:2018yfy}, which is maximally rigid.

%================================================================================================================
%================================================================================================================
% \newpage
%\bibliographystyle{physics}
%\bibliography{amplitude_refs}
%\end{document}
%\newpage
\providecommand{\href}[2]{#2}\begingroup\raggedright\endgroup

\end{document}